\documentclass[twoside,12pt]{article}

\usepackage{amssymb}
\usepackage{graphicx}
\usepackage{amsmath}
\usepackage{epsfig}

\title{
\vskip -70pt
\begin{flushright}
{\normalsize \ DAMTP-2004-122}\\
\end{flushright}
\vskip 25pt {\bf Beyond the Low Energy Approximation in Braneworld Cosmology} }
\vspace{1.4cm}
{\makeatletter
\author{Claudia de Rham {\thanks{e-mail address: C.deRham@damtp.cam.ac.uk}} \\
\small{\textsl{Department of Applied
    Mathematics and Theoretical Physics}}
\\ \small{\textsl{University of Cambridge}} \\
\small{\textsl{Wilberforce Road, Cambridge CB3 0WA, England}}}
\makeatother}
\date{\today}
\begin{document}
\maketitle
\vskip -10pt
\begin{abstract}
We develop a four-dimensional effective theory for
Randall-Sundrum models which
allows us to calculate long wavelength adiabatic
perturbations in a regime where the $\rho ^2$ terms characteristic
of braneworld cosmology are significant. This extends previous
work employing
the moduli space approximation.
We extend the treatment of the
system to include higher derivative corrections
present in the context of braneworld cosmology.
The developed formalism allows us to
study perturbations beyond the general long wavelength,
slow-velocity regime to which the usual moduli approximation is
restricted. It enables us to extend the study
to a wide range of braneworld cosmology
models for which the extra terms play a significant role.
As an example we
discuss high energy
inflation on the brane and analyze the key observational features
that distinguish braneworlds from ordinary inflation by considering
scalar and tensor perturbations as well as non-gaussianities.
We also compare inflation and Cyclic models and study how they can be
distinguished in terms of these corrections.
\end{abstract}

\maketitle


 \section{Introduction \label{introduction}}

Braneworlds offer a new approach to the phenomenology of both
cosmology and particle physics, and provide an alternative type of
low-energy string compactification. From the cosmologist's
perspective one of the most interesting aspects of these models is
that at high energies, such as those occurring in the early
universe, their gravitational behaviour is different from
conventional scalar-tensor cosmology. This is shown most clearly
in the well known ``modified Friedmann equation'', which contains
additional dark radiation and $\rho^2$ terms, where $\rho$ is the
energy density of matter on the brane.
In the simplest models,
at low energy
and at long wavelengths,
braneworlds behave like conventional scalar-tensor theories
\cite{GT,Mendes,KZ,SKT}.
However, in the early
Universe the low-energy condition
may well be violated.
Consequently, it makes sense to look beyond this conventional
limit. Unfortunately, this usually requires
a five-dimensional description, generically making analytic
solutions impossible, and requiring numerical methods. To get a
better insight into this regime it is useful to use approximation
methods that capture the essential physics, if not the
precise solution.
In this paper we
will consider approximation methods that allow us to go beyond
 the low-energy restriction.\bigskip

The Gauss-Codacci formalism relates the
five-dimensional Riemann tensor to its four-dimensional counterpart
induced on the brane. A modified four-dimensional
Einstein equation on the brane can be derived but the five-dimensional
nature of the model can not be completely ignored. Formally, the system
of equations obtained on the brane in terms of four-dimensional
quantities is not closed, as the modified Einstein equation contains
a new term: the electric part $E_{\mu \nu}$ of the five-dimensional Weyl
tensor. This term encodes information about the bulk geometry.
Assuming cosmological symmetry, the five-dimensional system can be
solved exactly. To start with, we use our knowledge of the
background
 Weyl tensor $E_{\mu \nu}$ to derive its behaviour for
adiabatic perturbations and solve the modified four-dimensional
Einstein equation. More precisely, for linear perturbations, we will neglect the
contribution from the part of the Weyl tensor $E_{\mu \nu}$ which
is transverse and vanishes in the background.
This procedure is consistent in an adiabatic regime only.\bigskip

Next, we extend this procedure by including some
  contribution
to the transverse part of
the Weyl tensor. We consider the contribution from a specific four-dimensional
tensor $A_{\mu \nu}$, that we check is
consistent with the proprieties of the Weyl tensor, and thus represents
  a natural candidate to consider. This tensor
  represents the term that would be derived if some
  four-dimensional Weyl-squared terms were included
in the four-dimensional effective action. Introducing a Weyl-squared
  term in the four-dimensional effective action preserves the
  conformal invariance of the effective theory and is therefore a
  natural term to consider \cite{paul}.

Around static branes, the first order perturbations on the brane
metrics have been solved exactly \cite{GT}.
It is well known that in the presence of an extra-dimension,
an infinite tower of Kaluza-Klein modes affects the four-dimensional geometry.
As long as the extra
dimension is of finite size, the modes are discrete. For static
branes, we show that
including the tensor $A_{\mu \nu}$ to the effective theory correctly
reproduces the effects of the first Kaluza-Klein mode in the
case where matter is present only on one brane. It seems therefore
natural to study the effect of the tensor $A_{\mu \nu}$ in more
general setups.

As a first example,
we use this formalism within the context of brane inflation for which
it reproduces
the nearly scale-invariant spectrum for density perturbations of
standard inflation. \bigskip

Recently,
the Ekpyrotic and Cyclic models
\cite{Ekp1,Ekp2,Cyclic1,Cyclic2,Cyclic3}
 have been introduced as an
alternative to inflation, giving a new picture to solve the
homogeneity, isotropy and flatness problems. It has
been shown that the Ekpyrotic and Cyclic models provide an alternative
scenario for the production of a nearly scale-invariant spectrum.
While the production of such a spectrum in
inflation is based on ``slow-roll'' conditions in an expanding
universe,
the Ekpyrotic and Cyclic
models, on the other hand, use a ``fast-roll'' potential in a
contracting universe for which the issue of the ``beginning of our Universe'' is
avoided. Further studies \cite{GKST,LPN} have shown an exact
duality
between the two
models in the production of density perturbations making the two
models hard to distinguish without bringing in results from the
observation of tensor perturbations. In this work we use our
prescription to compare the
behaviour of the fast-roll and slow-roll models in the case when $A_{\mu
  \nu}$ contributes
to the transverse part of Weyl tensor.
We examine how these
corrections
influence the
production of a scale-invariant spectrum, showing how they enable
us to distinguish between
general models with ``slow-roll'' and ``fast-roll''
conditions. Unfortunately, the
 difference between the standard inflation
scenario and the Cyclic model turns out to be negligible.  \bigskip


The paper is organised as follows. In section \ref{RSmodel}
we review the conventional Randall-Sundrum
model. At low-energy, we explain how a four-dimensional effective action
can be derived. The extension of the single brane
effective theory beyond the low-energy limit is discussed.
In section \ref{Sectioninflation}
we put this formalism
to use: we analyze a brane inflation model
in which the inflaton lives on the brane, and the typical energy scale
of inflation is not necessarily small compared to the brane tension.
The scalar and tensor perturbations are computed and we give an estimation
of non-gaussianity. We show how
 a model of brane inflation may always be reinterpreted as
an ordinary inflation model with a redefined potential, or
equivalently with redefined slow roll parameters.
In section \ref{§KK} we extend the previous model, taking
into account the possibility of the term $A_{\mu \nu}$ to be present
in the transverse
part of the Weyl tensor.
We also compare inflation and Cyclic models and analyze whether they can
be distinguished in terms of these corrections. Finally, we discuss
 the implications of our results in section
\ref{sectionconclusion}.

In appendix \ref{section p2 terms}, we derive the four-dimensional
effective theory for a general two brane Randall-Sundrum model which is
valid at large energy provided we work in the long wavelength
adiabatic regime.

We review the
exact behaviour of the  Kaluza-Klein modes around static branes
 in appendix \ref{section KK} and
show how in this limit the first mode can be reinterpreted in terms of local
four-dimensional quantities.  \bigskip


\section{Covariant Treatment of Randall-Sundrum Model \label{RSmodel}}

\subsection{Gauss-Codacci Equations}

In what follows, we shall be interested in the two
brane Randall-Sundrum model as a specific simple example of
braneworld cosmologies. We will see in which limit this model
may be extended to the one brane Randall-Sundrum model.
To begin with, let us
consider the two brane Randall-Sundrum model \cite{RS1,RS2} where
the spacetime is five-dimensional, with a compact extra dimension
having the topology of an $S_1 / \mathbf{Z_2}$ orbifold. The stress
energy of the bulk is assumed to be from a pure negative
cosmological constant $|\Lambda |= \frac{6}{\kappa\, L^2}$,
$\kappa=\frac{1}{M^3_5} \equiv \frac{L}{M^2_4}$, with $M_n$ the
$n$-dimensional Planck mass. There are two boundary branes (referred
as $\pm$-branes) 
located at
the fixed points of the $Z_2$ symmetry on which gauge and matter
fields are confined.

The Ricci tensor on each brane may be expressed in terms of
five-dimensional quantities by means of the Gauss-Codacci formalism
as in \cite{GCBinetruy99,SKT}:
\begin{eqnarray}
R_{\mu \nu }^{\text{(4d)}}=R_{\mu \nu }^{\text{(5d)}}+\left(
K_{\alpha \mu }K_{\nu }^{\alpha }-KK_{\mu \nu }\right) -E_{\mu \nu
}, \label{gauss eq}
\end{eqnarray}
where $E_{\mu \nu }$ is the electric part of the five-dimensional
Weyl tensor.
For a $Z_{2}$-symmetric brane, the extrinsic curvatures $K_{\mu \nu
}$ on each side of the brane are equal and opposite and can be
uniquely determined using the Isra\"{e}l matching condition
\cite{Israel}:
\begin{eqnarray}
\Delta K_{\mu \nu}=-\kappa\, \left(T^{\text{tot}}_{\mu
  \nu}-\frac{1}{3} g_{\mu \nu} T^{\text{tot}} \right),
\label{Israel}
\end{eqnarray}
where $g_{\mu \nu}$ is the induced metric on the brane and
$T^{\text{tot}}_{\mu \nu}$ is the total stress-energy on the brane
including the gauge and matter fields and the brane tension
contribution. We consider the tension on each brane to be fine-tuned
to their canonical value: $\mathcal{T}^{\pm}=\pm \frac{6}{\kappa\,
L}$ and we include some matter on each brane with stress-energy
tensor $T_{\mu \nu }^{(\pm )}$\ . Writing the extrinsic curvature in
terms of the stress-energy tensor, the projected Ricci tensor on
each brane can be expressed as:
\begin{eqnarray}
R_{\mu \nu }^{(\pm )}=\pm \frac{\kappa}{L} \left( T_{\mu \nu }^{(\pm
)}-\frac{1}{2}T^{(\pm )}g_{\mu \nu }^{(\pm )}\right)
-\frac{\kappa^2}{4}\left( T_{\mu }^{(\pm )\alpha }T_{\alpha \nu
}^{(\pm )}-\frac{1}{3}T^{(\pm )}T_{\mu \nu }^{(\pm )}\right) -E_{\mu
\nu }^{(\pm )}.  \label{Ricci1}
\end{eqnarray}
We can remark on two features in this modified Friedman equation. The
first one is the presence of terms quadratic in the stress-energy
tensor. The aim of our study will be to understand the implications of
those terms to braneworld cosmology. The second departure from the
standard four-dimensional Einstein equation arises from the presence
of the Weyl tensor $E_{\mu \nu}$ which is undetermined on the
brane. It is worth pointing out that when the
cosmological constant in the bulk is important, the length scale $L$ is
negligible compared to any other length scale present in the theory. 
In that case
the first term in (\ref{Ricci1}) dominates and we recover standard
four-dimensional gravity for the positive tension brane
 (up to the redefinition of the
four-dimensional Planck mass $\kappa \equiv \frac{L}{M^2_4}$).

\subsection{Exact Solution for Cosmological Symmetry}

Considering the five-dimensional Universe to be homogeneous and
isotropic in the three spatial directions, an exact static solution
of the five-dimensional Einstein equation can be found: the geometry
is of Schwarzschild-Anti-de-Sitter form with parameter
$\mathcal{C}$ associated with the black hole mass.

Knowing the bulk geometry exactly, the Weyl tensor can be
calculated. For homogeneous and isotropic metrics $E_{\mu \nu
}^{(\pm )}$ can be seen to have the form of the stress-energy tensor
for radiation \cite{KZ}:
\begin{eqnarray}
\begin{array}{l}
E_{\ \ 0}^{ (\pm )\; 0}=3 \frac{\mathcal{C}}{a^{4}_{\pm}}, \label{Euv+-} \\
E_{\ \ j}^{(\pm )\; i }=-\frac{\mathcal{C}}{a^{4}_{\pm}} \delta^i_j,
\end{array}
\end{eqnarray}
where the constant $\mathcal{C}$ is the same as the one
associated with the black hole mass.

Following the work done in
\cite{KZ,SMS,Binetruy99,Mukohyama,SSM,Mazumdar:2000sw,Binetruy01,
LangloisIntroduction,Langlois03},
the expression for the Weyl tensor can be introduced in the modified
Einstein equation (\ref{Ricci1}). Each of the two branes satisfy the
induced or modified Friedmann equation:
\begin{eqnarray}
H_{\pm}^2=-\frac{k}{a_{\pm}^2} \pm \frac{\kappa}{3 L}\rho_{\pm}
+\frac{\kappa ^2}{36 }\rho^2 _{\pm}+
\frac{\mathcal{C}}{a_{\pm}^4},
\label{modified friedman eq 1}
\end{eqnarray}
with the $+$ and $-$ indices designating the positive and negative
tension brane respectively, $H$ the induced Hubble parameter, $a$
the brane scale factor and $\rho$ the matter and radiation density
confined to each brane. In defining $\rho$ we have separated out
the part coming from the canonical tension of the brane.

\subsection{Four-dimensional Effective Action \label{BD}}
 In the
low-energy limit, assuming the matter and radiation
density on the branes to be much smaller than the magnitude of the
brane tensions, the $\rho^2 _{\pm}$ terms in (\ref{modified friedman
eq 1}) may be neglected  \cite{Mendes,KZ,BT,Shiromizu}. In that case, it has
been 
shown that for
the purpose of calculation of long wavelength perturbations, the
system may be well described by a four-dimensional effective theory
derived from the effective action of a scalar field minimally
coupled to gravity with non-minimally coupled matter:
\begin{eqnarray}
S=\int d^4 x\sqrt{-g} \left( \frac {L}{2 \kappa} R-\frac{1}{2}
\left(\partial \phi \right)^2 \right)+ S^-_m [g^-]+S^+_m [g^+],
\label{4daction}
\end{eqnarray}
with
\begin{eqnarray}
g^+ _{\mu \nu}=\left(\cosh \left(\phi / \sqrt{6} \right) \right)^2
g^{4d}_{\mu \nu},
\label{g+intermasofg4d} \ \ \
g^- _{\mu \nu}=\left(-\sinh \left(\phi / \sqrt{6} \right) \right)^2
g^{4d}_{\mu \nu},
\end{eqnarray}
and where $S^{\pm}_m [g^{\pm}]$ are the conventional four-dimensional
matter actions on each brane.
Here $g_{\mu\nu}^{\pm}$ are the induced geometries on each brane
which, in this approximation, are
seen to be conformal to each other. These equations are often written
in the conventional Brans-Dicke frame,
however we have chosen to use the more practical Einstein frame, at
the price of having non-minimally
coupled matter.
Assuming cosmological symmetry,
the equations of motion derived from this action are indeed
consistent with the modified Friedmann equations (\ref{modified
  friedman eq 1}) in the low-energy limit. Within this limit, the behaviour of long
wavelength, adiabatic linear perturbations may therefore be derived
from (\ref{4daction}), this has been done for instance in
\cite{PPLiddleLyth00,PPGen02}. However some braneworld models
consider situations where the $\rho ^2$ terms play a significant
role
\cite{chaoticinfationbrane,AAShtanov00,AALanglois00,CLL,PotLiddle01,
SDS,Maartens03,AAKoyama04}.
The purpose of the next section is therefore to understand how this
covariant formalism may be extended in order to satisfy the correct
density dependence in this more
general case. \bigskip

We stress that the extension considered will only be valid for {\it
adiabatic} perturbations where the stress energy of the matter on
each brane evolves adiabatically with the scalar field. However we
expect that our formalism may be used to model characteristically
five-dimensional effects in a more general setting of high energy or
velocity without directly dealing with the full five-dimensional
formalism as described in
\cite{5dLanglois00,5dCarsten00,5dDeffayet02,5dKoyama03}. \bigskip

\subsection{Treatment of the One Brane High-Energy Regime
 \label{secpureAdS}}
 In order to focus on
the effect of the quadratic terms in $T_{\mu \nu}$,
we will consider, in what follows, that the bulk has a pure
Anti-de-Sitter
 (AdS) geometry
in the background
($\mathcal{C}=0$). An extension to the case of matter in a
Schwarzschild-AdS bulk is presented in appendix \ref{section p2
terms}. In the background, when the bulk is fixed to pure AdS,
each brane evolves independently from each other. The negative
tension brane could be very close or
could be sent to infinity so as to recover the one brane
Randall-Sundrum model for the positive tension brane.
It is in this limit that we will consider the model in the
following.
Since we will be interested in quantities on the positive
tension brane only, the superscript $(+)$ will be suppressed.\bigskip

The Weyl tensor $E_{\mu \nu }$ in the modified Einstein equation
(\ref{Ricci1}) is a priori undetermined. This comes from the
five-dimensional nature of the theory and the fact that the system
of equations is not closed on the brane as the Weyl tensor mediates
some information from the bulk to the brane. However $E_{\mu \nu }$
is traceless \cite{SMS}, and we may decompose it into longitudinal
and tensor parts:
\begin{eqnarray}
&& E_{\mu \nu }=\mathcal{E}^{(L)}_{\mu \nu }+E_{\mu \nu }^{TT}, \label{decEuv }\\
&& \nabla_{\mu}E^{TT\, \mu}_{\,  \nu }=0, \ \ E^{TT\, \mu}_{\,  \mu }=0.
\end{eqnarray}
 Furthermore, since the Bianchi identity has to be satisfied for
the four-dimensional brane Einstein tensor, the divergence of the Weyl
tensor must satisfy:
\begin{align}
\nabla_{\mu}E^{\mu}_{ \nu }&= -\frac{\kappa^2}{4}\,
T^{\beta}_{\alpha}\left( \nabla_{\beta}(T^{\alpha}_{\nu}-\frac 1 3 T
\delta^{\alpha}_{\nu}) - \nabla_{\nu}(T^{\alpha}_{\beta}-\frac 1 3 T
\delta^{\alpha}_{\beta})
 \right), \label{divT2} \\
 &=\nabla_{\mu}\mathcal{E}^{(L)\, \mu}_{ \nu }.
\label{div Euv vect}
\end{align}
The longitudinal part may be determined up to a homogeneous tensor
which is absorbed in the transverse and traceless part $E_{\mu \nu
}^{TT}$. We may check that for any kind of matter satisfying
conservation of energy in a homogeneous and isotropic background,
the divergence in (\ref{divT2}) vanishes. This is consistent with
the fact that the longitudinal part of the Weyl tensor vanishes for
the background. (Even when the bulk geometry is not taken to be pure AdS
($\mathcal{C}\ne 0$), the part contributing to the Weyl tensor in
(\ref{Euv+-}) is homogeneous.) \\
This four-dimensional tensor part of the Weyl tensor is the only
quantity that remains unknown on the brane. For the purpose of the
first part of our study we will make the important assumption that
{\it ${\mathit E_{\mu \nu }^{TT}}$ can be neglected}. For pure AdS
background, the Weyl tensor vanishes and $ E_{\mu \nu
}^{TT}$ only comes in at first order in perturbation. \bigskip

%
%
 It is a well-known result that for long wavelength {\it
adiabatic}
 scalar perturbations, the
quantity $\frac{\delta
  \rho}{\dot{\rho}}$ is the same for any fluid with energy density
  $\rho$, satisfying conservation of energy,
  regardless of its equation of state \cite{LL}. In
particular, this is true for the conserved tensor $E_{\mu \nu}^{TT}$
(with density $\rho_{E}$ and $\omega_{E}=\frac{1}{3}$). The
transverse condition implies $\dot{\rho}_{E}=-4\,\frac{\dot a}{a} \rho_{E}$.
 We consider the fluid with energy density $\rho_{E}$ and
compare it with the energy density $\rho$ of any other fluid present
in the theory, or any scalar field $\Phi$ present in the theory. The
adiabaticity condition
 at long wavelengths imposes $\frac{\delta \rho_{E}}{\dot{ \rho_{E}}}
=\frac{\delta \rho}{\dot{
    \rho}}=\frac{\delta \Phi}{\dot{
    \Phi}}$ which requires that:
\begin{eqnarray}
\label{adiabaticargt} \delta \rho_{E}
=-4\,   \rho_{E} \, \frac{\dot a}{a}
\frac{\delta \rho}{\dot{ \rho}}
=-4\,   \rho_{E}\, \frac{\dot a}{a} \frac{\delta \Phi}{\dot{ \Phi}}.
\end{eqnarray}
Since the tensor $E_{\mu \nu }^{TT}$ vanishes in the background,
 for adiabatic perturbations,
its
contribution
vanishes as well, $\delta
\rho_{E}=0$. Thus in the adiabatic limit
the long wavelength scalar perturbations of
$E_{\mu \nu }^{TT}$ can be neglected. \\
In other words,
in an adiabatic regime, the perturbations follow the same evolution as
a general background solution.
Having a non-vanishing perturbed $E_{\mu
\nu }^{TT}$ could be seen as the introduction of a black hole at the
perturbed level, changing the background from pure AdS to
Schwarzschild-AdS. For the one brane limit, this is not compatible
with the bulk boundary
conditions at infinity as the bulk geometry would diverge
exponentially.  \bigskip

For the
three-dimensional-tensor perturbations, the same approximation will
be made, although no analogous argument may be given.
It is in the context of this approximation that we shall
consider inflation on the brane in the next section. \bigskip

In appendix \ref{section p2 terms} a slightly more elaborate version
of the treatment of the Weyl tensor is given for the two
brane scenario in a general Schwarzschild-AdS background. However
as stated earlier we shall for simplicity focus on the one brane
limit. In the next section we use this approach to study inflation
on the positive tension brane.


\section{Inflation on the brane \label{Sectioninflation}}
In this section we shall consider the inflaton to be an additional
scalar field living on the positive tension brane. If the energy
scale of inflation is well below the brane tension then this system
is well described by the Brans-Dicke theory as in section
\ref{BD}. However, if the typical energy scale of inflation is
comparable to or larger than the brane tension then we may use the
formalism of the previous section to get a better insight.
\subsection{Background}
For simplicity, let us assume that the brane is spatially flat in
the background ($k=0$). The stress-energy tensor of the inflaton
is given by:
\begin{eqnarray}
T_{\mu \nu }=\partial _{\mu }\varphi \partial _{\nu
}\varphi -\left( \frac{1}{2}\left( \partial \varphi \right) ^{2}+V(\varphi
)\right) \,g_{\mu \nu }.
\end{eqnarray}
The inflaton $\varphi$ evolves in a potential $V(\varphi)$ which is
assumed to satisfy some slow-roll conditions which will be specified
later. The energy density is:
\begin{eqnarray}
\rho=-T_0^0= V(\varphi_0)+ \frac{1}{2a^{2}}\dot{\varphi}_{0}^{2}.
\label{modified friedman 2}
\end{eqnarray}
where a dot
represents derivative with respect to the conformal time.

When the kinetic energy of the scalar field is assumed to be
negligible compared to its potential energy,
$\frac{1}{a^{2}}\dot{\varphi}_{0}^{2} \ll  V(\varphi_0)$, the
modified Friedmann equation (\ref{modified friedman eq 1}) reads:
\begin{eqnarray}
H^2&\simeq&\frac{2}{L^2 \mathcal{T}}V\left(1+\frac{V}{2\mathcal{T}}
\right)\simeq \text{constant}. \label{modified friedman 3}
\end{eqnarray}
It is worth pointing out that an expansion in
$\rho/ \mathcal{T}$ is equivalent to an expansion in $L^2 H^2$.
Usually in the low-energy limit
 $L^2H^2 \ll 1$ and $2 \rho/ \mathcal{T} \sim  L^2H^2$.
In what follows we will keep all terms in
$L^2H^2$ so that the theory remains valid at high energy, when
$L^2H^2\gg 1$.
As mentioned before, in the limit where the length scale $L$ is
negligible compared to the other length scales of the theory, we should
recover the standard four-dimensional results. \\
Since the inflaton is confined on the four-dimensional brane, the
Klein-Gordon field equation of motion coming from the conservation
law of the stress-energy tensor $\nabla _{\mu }T_{\nu }^{ \mu }=0$
is the usual four-dimensional equation:
\begin{eqnarray}
V_{,\varphi}(\varphi _{0})=-\frac{1}{a^{2}}\,\ddot{\varphi}_{0}-2\,\frac{
\dot{a}}{a^{3}}\,\dot{\varphi}_{0}.
\end{eqnarray}
This simplifies to the conventional result in the slow-roll regime $ \frac{d}{d
\tau}\frac{\dot{\varphi}_{0}}{a}
 \ll H \dot{\varphi}_{0}$:
\begin{eqnarray}
\dot{\varphi}_{0}\simeq -\frac{a V_{,\varphi}}{3H}.
\end{eqnarray}
 We may now use the formalism of the
 previous section to consider scalar and tensor perturbations as
 well as an estimation of non-gaussianity.
%
%
\subsection{Linear Scalar Perturbations}
We consider linear isotropic perturbations around this conformally
flat background. In longitudinal gauge,
\begin{eqnarray}
&& ds^2=a^2(\tau)\, \left(-1+2 \Phi\right)
d \tau^2+a^2(\tau)\,\left(1+2\Psi
\right)d\mathbf{x}^2 ,\\
 &&\varphi (\tau, \mathbf{x})=\varphi _{0}(\tau)+\delta \varphi(\tau, \mathbf{x}).
\end{eqnarray}
As mentioned previously, in order for the equation of motion to be
consistent with the Bianchi identity the Weyl tensor must include a
term cancelling the divergence of the quadratic terms in $T_{\mu
\nu}$. Using the background equation of motion, we need to have:
\begin{align}
\nabla_{\mu}E^{\mu}_{ \nu }&= -\frac{\kappa ^2}{4}\,
T^{\beta}_{\alpha}\left(
\nabla_{\beta}(T^{\alpha}_{\nu}-\frac 1 3 T
\delta^{\alpha}_{\nu}) - \nabla_{\nu}(T^{\alpha}_{\beta}-\frac 1 3 T
\delta^{\alpha}_{\beta})
 \right), \\
&=-\frac{\kappa^2}{4}\left(
\begin{array}{c}
0 \\
-\frac{2}{3}\frac{\dot{\varphi}_{0}^{2}}{a^{4}}\left(
-\ddot{\varphi} _{0}\,\delta \varphi
+\frac{\dot{a}}{a}\dot{\varphi}_{0}\,\delta \varphi +
\dot{\varphi}_{0}^{2}\,\Phi +\dot{\varphi}_{0}\,\delta
\dot{\varphi}\right) _{,i}
\end{array}
\right).
\end{align}
We stress that the time-like component of this divergence
vanishes.
This is a particularity of the stress-energy coming from a scalar
field and will not, as far as we know, be true for a general fluid.
This allows us to decompose the Weyl tensor in the simple
form:
\begin{eqnarray}
& E_{\mu \nu }=\mathcal{E}_{\mu \nu }^{(L)}+E_{\mu \nu }^{TT}
\label{Euv1}, \\
\text{with } & \ \  \mathcal{E}_{\mu \nu }^{(L)} =\left(
\begin{array}{cc}
0 & 0 \\
0 & X_{,ij}-\frac{1}{3}\nabla ^{2}X\gamma _{ij}
\end{array}
\right),\label{Euv2} \ \ \
\nabla_{\mu}\mathcal{E}^{(L)\ \mu}_{ \nu }=
\left(
\begin{array}{c}
0 \\ \frac{2}{3}\nabla ^{2}X_{,i}
\end{array}
\right).
\end{eqnarray}
This is remarkable since an expression for the longitudinal part of
the Weyl tensor has been found without needing to solve a
differential equation involving time derivatives, which would have
required us to specify some initial conditions. The expression for $X$ is
given by:
\begin{eqnarray}
\nabla ^{2}X &=&
\frac{\kappa^2}{4}\frac{\dot{\varphi}_{0}^{2}}{a^{2} }\left(
-\ddot{\varphi}_{0}\,\delta \varphi +\frac{\dot{a}}{a}\dot{\varphi}
_{0}\,\delta \varphi +\dot{\varphi}_{0}^{2}\,\Phi
+\dot{\varphi}_{0}\,\delta \dot{\varphi}\right). \label{nablaX}
\end{eqnarray}
Having an expression for the Weyl tensor (\ref{Euv2}), we need to
solve the modified Einstein equation (\ref{Ricci1}):
\begin{eqnarray}
\begin{array}{l}
G_{\mu \nu}=\frac{6}{\mathcal{T}L^2}T_{\mu
\nu}-\frac{9}{\mathcal{T}^2L^2} \left(T_{\mu}^{\alpha}T_{\alpha
\nu}-\frac{1}{3}T T_{\mu \nu}-\frac{1}{2}T_{\alpha \beta}T^{\alpha
\beta}g_{\mu \nu}+\frac{1}{6}T^2g_{\mu \nu} \right)
-\mathcal{E}_{\mu
\nu}^{(L)} 
\end{array}\label{modEq}
\end{eqnarray}
When we derive the $(ij)$ (with
$i\ne j$) component of this equation, a second interesting feature arises
from the ansatz (\ref{Euv2}). Indeed the $(ij)$-equation
directly points to the presence of an effective {\it
anisotropic} stress $X$ at high energy:
\begin{eqnarray}
X &=& \Psi - \Phi . \label{eqX}
\end{eqnarray}
Since only the $\rho ^2$ terms contribute to $X$, (eq.(\ref{nablaX})
is quartic in the fields so quadratic in the energy density), we
notice that at high-energy, the $\rho ^2$ terms play the role of an
effective anisotropic stress.

For linear perturbations, the $(0i)$-component of the modified
Einstein equation (\ref{modEq}) reads:
\begin{eqnarray}
\delta \varphi =-\frac{2L}{\kappa\,\dot{\varphi}_{0}\sqrt{
1+L^{2}H^{2}}}\left( \frac{\dot{a}}{a}\Phi +\dot{\Psi}\right).
\label{deltaphi}
\end{eqnarray}
Using this expression in (\ref{nablaX}) and (\ref{eqX}), we can
express $\dot{\Phi}$ in terms of $\Phi$ and $\Psi, \dot{\Psi},
\ddot{\Psi}$. Using this, as well as the
expression for $\delta \varphi$, in the $(00)$-component of the
modified Einstein equation, we find the following relation between
$\Phi$ and $\Psi$:
\begin{eqnarray}
\Phi=\left(1- \frac{L^2a\dot{H}}{1+L^2H^2}\right)\, \Psi= \left(1-
\frac{a}{H} \frac{d}{d \tau}\left( \ln \sqrt{1+L^2H^2}\right) \right)\, \Psi .
\end{eqnarray}
Note that the first and second time derivative of $\Psi$
cancel exactly,
 giving a
surprisingly simple expression for the anisotropic stress. In the
low-energy regime where $L^2H^2 \ll 1$, the anisotropic part cancels
out, and the usual result is recovered.

Substituting this relation between $\Phi$ and $\Psi$ into the previous
expression we had for $\dot{\Phi}$ in terms of $\Phi, \Psi, \dot{\Psi},
\ddot{\Psi}$ gives the decoupled second order equation for $\Psi $:
\begin{eqnarray}
\begin{array}{l}
0= \ddot{\Psi}-\nabla ^{2}\Psi +2\left(
\frac{\dot{a}}{a}-\frac{\ddot{
\varphi}_{0}}{\dot{\varphi}_{0}}\right) \left(
\dot{\Psi}+\frac{\dot{a}}{a}
\Psi \right)  \\
\phantom{0= } +2\,a\dot{H} \left( \Psi-\frac{L^2}{1+L^2H^2}\left( H^2 \Psi
-H\frac{\ddot{\varphi}_0}{a \dot{\varphi}_0}\Psi
+\frac{H}{a}\dot{\Psi} \right)
 \right) \\
\phantom{0= } -L^2 \dot{H}^2\frac{2-L^2H^2}{1+L^2H^2}\Psi
-\frac{L^2\ddot{H}H}{1+L^2H^2}\Psi.
\end{array}
\label{eqPsi}
\end{eqnarray}
We may point out that the only difference with usual four-dimensional
scalar perturbations will arise from this second-order equation for
$\Psi $. We have indeed already mentioned that for a given geometry,
the Klein-Gordon equation for the scalar field remains the usual
four-dimensional one $\nabla _{\mu }T_{\nu }^{\mu }=0$:
\begin{eqnarray}
\delta \ddot{\varphi}=\nabla ^{2}\,\delta \varphi -2a^{2}\Phi
V_{,\varphi }+a^{2}\delta \varphi V_{,\varphi \varphi
}-\dot{\varphi}_{0}\left(3\dot{\Psi}+\dot{\Phi}\right)-2
\frac{\dot{a}}{a}\delta \dot{\varphi}.
\end{eqnarray}
We can define the gauge invariant variable $u$ related to the metric
perturbation by $u=z\Psi$ with $z=\frac{a}{\dot{\varphi}_0\,
\sqrt{1+L^2H^2}}$. It is more significant to express it in terms of
the scalar field perturbations:
\begin{eqnarray}
\delta \varphi=\frac{-2 L}{a\, \kappa}\left(
\dot{u}+\frac{\ddot{\varphi}_0}{\dot{\varphi}_0}u\right),
\end{eqnarray}
ie. precisely the conventional relation between the Mukhanov
variable $u$ and the scalar field perturbations  \cite{Mukhanov}
(after defining the
four-dimensional Planck mass to be $M^2_4=\frac{L}{\kappa}$).

Using the equation for $\Psi$, the Mukhanov variable $u$ satisfies
the second order differential equation:
\begin{eqnarray}
 \ddot{u}_{k}+\left(k^2-\frac{\beta}{\tau^2}
\right)u_{k}=0 \label{eqU1}, \ \ \ \text{with } \
\frac{\beta}{\tau^2}\equiv\frac{\ddot{\theta}}{\theta}, \ \
 \theta=\frac{H}{\dot{\varphi}_0}, \label{equ}
\end{eqnarray}
which is precisely the standard equation.
The braneworld modifications have only altered the expression for
the Hubble 'constant' $H$ in the modified Friedmann equation. The
corrections to the observable quantities at the linear perturbation
level
directly result from this background modification alone.\bigskip
%

We may now follow the prescription of \cite{GKST}.  Assuming that
$\beta$ may be treated as a constant \textendash the conditions for this
assumption will be specified by the slow-roll parameters later on \textendash \,
an analytical solution of (\ref{equ}) for the k-modes $u_k$ can be
found. The integration constant may be fixed by requiring the scalar
field fluctuations to be in the Minkowski vacuum well inside the
Hubble radius, and by normalizing the modes as:
\begin{align}
& k^2  \tau^2 \gg \beta \sim 3 \epsilon-\eta, \notag \\
& u_k \sim \frac{i e^{-i k \tau}}{(2 k)^{3/2}}, \\
& \delta \varphi_k \sim -\frac{L}{a \kappa}\frac{e^{-i k \tau}}{\sqrt{2
    k}}\left(1-\frac{i}{k \tau} \right),
\end{align}
which corresponds to the Bunch-Davis vacuum \cite{BD}.\\
In the long wavelength regime, the modes satisfy:
\begin{align}
& 
k^2 \tau^2 \ll 1,  \notag \\
&
u_k
\approx \frac{\sqrt{\pi}\, k^{-3/2}}{2^{3/2}\sin(n\pi)\Gamma (1-n)}
\left(\frac{-k\tau}{2} \right)^{\frac{1}{2}-n}
\begin{array}{r}
\left(1-e^{-i \pi n}\frac{\Gamma (1-n)}{\Gamma (1+n)}
\left(\frac{-k\tau}{2} \right)^{2n} \right. \\
\left. -\frac{\Gamma (1-n)}{\Gamma
(2-n)} \left(\frac{-k\tau}{2} \right)^{2}\right)
\end{array}
,  \label{eqpsik}
\end{align}
with the index $n=\sqrt{\beta+\frac{1}{4}}$. \bigskip

In order to have a physical interpretation of this result, we can
relate it to the gauge invariant curvature perturbation on {\it
comoving} hypersurfaces
 $\zeta_{\varphi}=\Psi-\frac{\dot a}{a}\frac{\delta
 \varphi}{\dot{\varphi}}$.  To be entirely rigorous,
 we can also
 consider the curvature perturbations on
uniform-energy-density hypersurfaces $\zeta_{\rho}$ or the curvature
perturbations on uniform-{\it effective}-energy-density
hypersurfaces $\zeta_{\rho_{eff}}$:
\begin{eqnarray}
\zeta_{\varphi}=\Psi-\frac{\dot a}{a}\frac{\delta
 \varphi}{\dot{\varphi}}, \ \
\zeta_{\rho}=\Psi-\frac{\dot a}{a}\frac{\delta \rho}{\dot{\rho}} \ \
\text{or} \ \ \zeta_{\rho_{eff}}=\Psi-\frac{\dot a}{a}\frac{\delta
\rho_{\text{eff}}}{\dot{\rho_{\text{eff}}}},
\end{eqnarray}
where $T_{\mu \nu}^{\text{eff}}$ is given by the right hand side of
(\ref{modEq}).
All those three quantities are conserved at long wavelengths by
conservation of energy. We may indeed check that for energy density
coming from an inflationary scalar field,
\begin{eqnarray}
\zeta_{\rho}&=&\Psi+\frac{a^2\,\delta \rho_{\varphi}}{3\dot{\varphi}_0^2}, \notag \\
\dot{\zeta}_{\rho} &=&-\frac{4
  \frac{\dot a}{a}+2\frac{\ddot{\varphi}_0}{\dot{\varphi}_0}}{3
    \dot{H}}
\, \nabla ^{2}\Psi.
\end{eqnarray}
Since $T_{\mu \nu}^{\text{eff}}$ is also conserved 
(despite not being linearly related to
  $T_{\mu \nu}$), we may again argue that $\zeta_{\rho_{eff}}$ is conserved at long
  wavelengths. As pointed out in section \ref{secpureAdS}, if the
 adiabaticity condition holds for linear
perturbations, then $\zeta_{\rho_{eff}}$, $\zeta_{\rho}$ and
$\zeta_{\varphi}$ should coincide at long wavelengths,
\begin{eqnarray}
\text{for } k^2 \tau^2 \ll 1,  \ \ \ \  \langle \zeta_{\rho}^2 \rangle
\simeq
\langle \zeta_{\rho_{eff}}^2
\rangle \simeq  \langle \zeta_{\varphi}^2 \rangle. \notag
\end{eqnarray}
In the following we will concentrate on $\zeta_{\varphi}$. In the
slow-roll regime, at long wavelengths,
\begin{eqnarray}
\delta \varphi_k \simeq -\frac{2L H}{\kappa}u \simeq -\frac{i L H}{\sqrt 2
\kappa k^{3/2}}\, e^{-i k \tau}, \\
\zeta_{\varphi} \simeq -\frac{\dot a}{a}\frac{\delta
\varphi_k}{\dot{\varphi}_0} \sim \frac{i a }{\sqrt 2\,
k^{3/2}}\frac{H^2}{\dot \varphi_0}\, e^{-i k \tau}.
\end{eqnarray}
The power spectrum is therefore given by the standard expression:
\begin{eqnarray}
\mathcal{P} &\sim & k^3  \langle \zeta ^2 \rangle \sim \left.
\frac{L^2}{2 \kappa^2 }\frac{a^2 H^4}{\dot \varphi_0^2}
\right|_{\tau=\tau ^*} \sim \left.
\frac{H^6}{V_{,\varphi}^2} \right|_{\tau=\tau ^*} \\
&\sim & \left.
\frac{V^3(1+V/2\mathcal{T})^3}{M^6_4\, V_{,\varphi}^2} \right|_{\tau=\tau ^*},
\end{eqnarray}
with $\tau ^{*}$ the time of horizon crossing when $k=aH$. Once
again the only departure from the standard four-dimensional
inflation result only comes from the modification of the
Friedmann equation (\ref{modified
  friedman 3})
at the
background level. Expressed in terms of the potential, the power
spectrum will therefore get an overall factor of $\left(1+\frac{V}{2
  \mathcal{T}}
\right)^3$, as mentioned in \cite{chaoticinfationbrane}. For a given
potential,
it will appear to be redder than for chaotic inflation.\bigskip

For {\it small $\beta$} (we will study in the following the
conditions for $\beta$ to be small), the spectral index is given by:
\begin{eqnarray}
n_{S}-1&=&\frac{d \ \ln \mathcal{P}_{\zeta}}{d\ \ln
  k}=
 -2\beta+2\beta^2+\mathcal{O}(\beta^3).
\end{eqnarray}
We define the slow-roll parameters:
\begin{eqnarray}
\epsilon
=-\frac{\dot{H}}{a H^2} \ \text{ and } \label{epsilon} \
\lambda^{(n)}=\frac{d^{n}\ln \epsilon}{ d \ln a\ ^n}.
\end{eqnarray}
Each parameter $\lambda^{(n)}$ may be treated as a constant if
$\lambda^{(n+1)}\ll 1$ (writing $\lambda^{(0)}=\epsilon$).
In terms of those parameters, $\beta$ takes the {\it exact } form:
\begin{eqnarray}
\beta=-\frac{a^2H^2\tau^2}{4}
\left[ 
\begin{array}{l}
2\lambda^{(2)}-\left(2+ \lambda^{(1)}\right)\left(2\epsilon+
\lambda^{(1)}\right) \\ 
 -\frac{L^2H^2}{(1+L^2H^2)^2}\ \epsilon
\left( 2+6\epsilon +L^2H^2(2+3\epsilon)
 \right)
 \end{array}
   \right]
. \label{beta2}
\end{eqnarray}
We have made two assumptions on $\beta$: we
required it to be small, $\beta \ll 1$, and we treated it as a constant.
For $\beta$ to be small, the parameters $\epsilon$ and
$\lambda^{(1)}, \lambda^{(2)}$ need to be small. This will be
translated into the slow-roll parameter condition. Neglecting the
$\lambda^{(n)}$, ($\epsilon$ can be considered as a constant),
$\epsilon=-\frac{\dot{H}}{a H^2}$ implies that $a \simeq -\frac{1}{H
\tau^{\frac{1}{1-\epsilon}}}$. The overall coefficient in the
expression (\ref{beta2}) of $\beta$ is therefore
$a^2H^2\tau^2 \sim \tau ^{-2\frac{\epsilon}{1-\epsilon}}$.
Thus it is consistent to treat
$\beta$ as a constant as has been done to get the expression (\ref{eqpsik})
in the regime where
$\epsilon \ll 1,\ \lambda^{(1)} \ll 1$ and $\lambda^{(2)} \ll 1$.

Up to first order in the slow-roll parameters $\epsilon$ and
$\lambda^{(1)}$, (considering all other $\lambda^{(n)}$ to be negligible),
the spectral index takes the form:
\begin{eqnarray}
n_{S}-1&=&-2\epsilon -\lambda^{(1)}-\frac{L^2H^2}{1+L^2H^2}\ \epsilon .
\end{eqnarray}
We recover the standard result for the spectral index if we define the
second slow-roll parameter $\eta$ such that:
\begin{eqnarray}
n_{S}-1&=& -6\epsilon+2 \eta+\mathcal{O}(\epsilon, \eta).
\end{eqnarray}
In that case, the parameters $\epsilon$ and $\eta$ may
be expressed in terms of the scalar field as pointed out in
\cite{chaoticinfationbrane,BraneSlowRoll03,Calcagni:2003sg}:
\begin{eqnarray}
&& \epsilon=-\frac{\dot{H}}{a H^2}= \frac{L}{2\, \kappa
}\frac{1+V/\mathcal{T}}{(1+V/2\mathcal{T})^2}\frac{V_{,\varphi}^2}{V^2}
,\label{epsilon1}\\
&& \eta
=-\frac{(HH')' }{H^2H'}-\frac{1}{2}\frac{L^2H^2}{1+L^2H^2}\, \epsilon
=\frac{L}{\kappa}\frac{1}{1+V/2 \mathcal{T}}\frac{V_{,\varphi
\varphi}}{V} \label{eta},
\end{eqnarray}
where a prime represents the derivative  $\frac{d}{d t}$ 
with respect to the proper time $t$: $\frac{d}{d t}=\frac{d}{a
  d\tau}$ .

If we worked at higher order in the slow-roll parameters, there would
 again be some departure from standard four-dimensional gravity in the
spectral index that could be eliminated by an adequate redefinition of the third
slow-roll parameter $\xi^2$ (as defined for instance in 
\cite{LL, langlois,Calcagni:2003sg}).
We do not wish to show the calculation explicitly here, but it can be
seen that if the second order parameter $\xi^2$ is defined in our case such that
$\xi^2=\frac{1}{1+V/2\mathcal{T}}\xi_{\text{4d}}^2+\frac{2\,
L^2 H^2}{1+L^2H^2}\, \epsilon ^2$, the power spectrum to second order
in the slow-roll parameters will recover the same form as in the
standard four-dimensional chaotic inflation. \bigskip

At this point, from the
knowledge of the amplitude of the scalar perturbations {\it at long
wavelengths} and their scale-dependence, it is not possible to
distinguish between a model of standard four-dimensional chaotic
inflation with potential $V^{(\text{4d})}$ satisfying the standard
slow-roll conditions with parameters $\epsilon_{\text{4d}}$,
$\eta_{\text{4d}}$, $\xi_{\text{4d}}$, and a model of brane inflation
with a potential $V$, such that at the beginning of inflation when
the first modes exit the horizon,
\begin{eqnarray}
V^{(\text{4d})}&\simeq&V\left(1+\frac{V}{2\mathcal{T}}
\right)\left(1+\frac{V}{\mathcal{T}} \right), \\
V_{,\varphi}^{(\text{4d})}&\simeq&V_{,\varphi}\,
\left(1+\frac{V}{\mathcal{T}} \right), \\
V_{,\varphi \varphi }^{(\text{4d})}&\simeq&
V_{, \varphi \varphi}\,
\left(1+\frac{V}{\mathcal{T}} \right), \\
V_{, \varphi \varphi \varphi}^{(\text{4d})}&\simeq&
V_{, \varphi \varphi \varphi}\,\left(1+\frac{V}{\mathcal{T}}\right)
+ \frac{V_{,\varphi}^3}{V^2}\frac{\left(1+V/\mathcal{T}
  \right)^3}{\left(1+V/2\mathcal{T}
  \right)}\frac{4V/\mathcal{T}}{1+2V/\mathcal{T}\left(1+V/2\mathcal{T}
  \right)}\, . 
\end{eqnarray}
We argue that observations of long wavelength scalar perturbations
alone are not enough to differentiate between standard chaotic
inflation and inflation on a brane with a potential satisfying the
modified slow-roll conditions $\epsilon, \eta\ll 1$, $\eta$
as given in (\ref{eta}). Such observations are not sufficient
to distinguish between inflation occurring in a purely
four-dimensional universe and on a brane embedded in a fifth
dimension. 

To extend this study we will consider, in the next
section, typical effects that may arise on the brane due to the
non-local nature of the theory. We will see how the behaviour of
the perturbations may differ in that case from the
standard case. However we may first notice that in the limit of
large wavelengths comparing scalar and tensor perturbations will give
a different signature in steep brane inflation than in standard
four-dimensional inflation with one scalar field.
%
%
%
%
\subsection{Tensor Modes \label{section tensor1}}
The scalar field $\varphi$ being the only source of matter of the
theory, the effect of the quadratic term in the matter stress-energy
tensor on the behaviour of tensor perturbations will strictly be a
``background effect'' (ie. introducing some $T_{\mu
\alpha}^{0}T_{\nu
  \beta}^{0}h^{(t)\, \alpha \beta}$-kind of source terms, with $T_{\mu
  \alpha}^{0}$ the background value of the stress-energy
tensor). For purely tensor perturbations, the vector part
(\ref{decEuv }) of the Weyl tensor must vanish.
Indeed, we can consider the metric perturbation:
\begin{eqnarray}
ds^2=-a^2(\tau)\, d\tau^2+a^2(\tau)\,\left(\delta_{ij}+h_{ij}
\right)dx^idx^j,
\end{eqnarray}
where the three-dimensional tensor $h_{ij}$ is transverse and
traceless $h^{i}_{i}=0,\ \partial_{i}h^i_j=0$, and indices are
raised with $\delta^{ij}$.
With respect to this metric, we may indeed check
that $\nabla_{\mu}E^{\mu}_{ \nu }=0$.
Using the background
equation of motion, the tensor modes satisfy the standard equation:
\begin{eqnarray}
\ddot{h}_{ij}+2\frac{\dot a}{a}\dot{h}_{ij}-\nabla ^2 h_{ij}=0,
\label{eqh}
\end{eqnarray}
where the only difference to the chaotic inflation case
arises in the relation of the scale factor $a$ to the potential
through the modified Friedmann equation (\ref{modified friedman 2},
\ref{modified friedman 3}).
 We can therefore
treat the tensor perturbations the same way as is usually done in
standard inflation. The power spectrum is given by
\cite{chaoticinfationbrane,langlois}:
\begin{eqnarray}
\mathcal{P}_g&=& \frac{72}{M_4^2}H^2|_{\tau=\tau^*}\\
&\simeq&  \left. \frac{24}{M_4^4}V\left(1+\frac{V}{2\mathcal{T}}
\right)\right|_{\tau=\tau^*} \label{Pg}.
\end{eqnarray}
Here again, we notice that for a given potential, the power
spectrum for gravitational waves will appear to be redder. However the
overall factor is less important than for the scalar power spectrum
which will be reflected in the ratio.

The tensor spectral index can be derived the usual way:
\begin{eqnarray}
n_T=\frac{d \ln \mathcal{P}_g}{d \ln
  k}=
-\frac{L^2 \mathcal{T}}{6}\frac{V_{,\varphi}^2}{V^2}
\frac{1+\frac{V}{\mathcal{T}}}{(1+\frac{V}{2\mathcal{T}})^2}
=-2 \epsilon,
\end{eqnarray}
and the ratio $r$ of the amplitude of the tensor perturbations
at Hubble crossing to the scalar perturbations is modified by a factor:
\begin{eqnarray}
r &\simeq & \left. \frac{\epsilon }{1+V/\mathcal{T}}\, \right|_{\tau
   =\tau ^*}.
\end{eqnarray}
As mentioned in \cite{chaoticinfationbrane}, even though the tensor
spectral tilt remains the same as in standard inflation: $n_T=-2
\epsilon$,
for high-energy $V/\mathcal{T} \gg 1$, the ratio of tensor to scalar
perturbations on the brane will in general be smaller than what is
expected from ordinary single scalar field inflation.
However for slightly more complicated models of four-dimensional
inflation, such as hybrid inflation, the ratio $r$ will similarly be
reduced in comparison to chaotic inflation and the relation of tensor
to scalar perturbations is still not enough to distinguish between
such a model and a scenario of brane inflation. The Cyclic-model
predicts as well a low amplitude for the tensor modes. In the
following section we shall study some more fundamental features of
the five-dimensional nature of this model.
%
%
\subsection{Estimation of the Non-Gaussianity}
In order to have a consistency check on the validity of the
perturbative approach, we estimate the non-gaussianity
corrections to the power spectrum. We consider the non-gaussian part
to be strongly dominated by the scalar field perturbations. In
general, an estimate can be given by comparing the cubic terms
in the Lagrangian to the quadratic ones. Unfortunately this method
cannot be used here, so as a first approximation we will compare the
quadratic term in the redefined gauge invariant comoving curvature
to the previous quantity. Using the result of
\cite{Maldacena,ABMR}, to second order, the gauge invariant
curvature perturbation on uniform-energy-density hypersurfaces
$\zeta^{(2)}_{\varphi}$ is given by:
\begin{eqnarray}
\zeta^{(2)}_{\varphi}&=&\frac{\left(\dot \Psi +2\frac{\dot a}{a}\Psi
+\frac{\dot a}{a}\frac{\delta \dot{ \varphi}}{\dot \varphi}
 \right)^2}{\frac{\ddot a}{a}+\frac{\dot a^2}{a^2}-\frac{\dot a}{a}
\frac{\ddot \varphi}{\dot \varphi}}.
\end{eqnarray}
Again we could be interested in the curvature perturbation on
uniform-effective-energy-density hypersurfaces or on comoving
hypersurfaces instead, but for the purpose of this estimate, those
three quantities will give a coinciding result. The ratio of the
second term to the first one in this expansion gives an estimation
which is the same as in the context of slow-roll inflation:
\begin{eqnarray}
\frac{\sqrt{\zeta^{(2)}_{\varphi}}}{\zeta^{(1)}_{\varphi}}\simeq
\frac{3}{\sqrt{2}}\,  \epsilon.
\end{eqnarray}
In this estimate, some explicit $L^2H^2$ corrections will arise at
second order in the slow-roll parameter $\epsilon$. However this remains an
estimate and it is presented here only to point out that
the non-linear terms seem to be damped in comparison to the
linear ones by an order of magnitude proportional to the slow-roll
parameter which is precisely what is found in the context of standard
four-dimensional inflation.


\section{Corrections from the Weyl Tensor \label{§KK}}

In section \ref{secpureAdS}, we have pointed out the presence of a  
transverse and traceless term in the Weyl tensor
$E^{TT}_{\mu  \nu}$. So far, this term has been neglected. This was
motivated by the fact that $E^{TT}_{\mu  \nu}$ vanishes for the background
 and we can check that within this approximation, we recover the long
 wavelength limit of the exact five-dimensional theory. 
We gave as well an argument to justify
this approximation in the
regime of long wavelength adiabatic perturbations for scalar
fields. However, there is no reason for the tensor $E^{TT}_{\mu  \nu}$
to remain negligible in general. This term actually encodes
information about the bulk geometry. At the perturbed level, the
bulk geometry is not purely AdS anymore. The fluctuations in the bulk
geometry will generate some Kaluza-Klein (KK) corrections on the
branes. Those KK modes are mediated by the only term which 
remains undetermined from a purely four-dimensional point of view:
the 
tensor $E^{TT}_{\mu  \nu}$. 

In this section, we want to modify the four-dimensional effective theory in
order to study the typical kind of corrections that may arise from the
perturbed bulk geometry. The only modification
 consistent with the overall five-dimensional nature of
the model is to add a contribution coming from $E^{TT}_{\mu  \nu}$.
We want to modify the effective theory in
order to accommodate terms that are negligible in the long wavelength
limit of the five-dimensional theory. We therefore need to consider
terms 
of higher order in derivatives (compared to the other terms already present in
the theory).
From the properties of the
five-dimensional Weyl tensor, together with the Bianchi identity on
the brane, $E^{TT}_{\mu  \nu}$ satisfies the following properties:
\begin{itemize}
    \item $E^{TT}_{\mu  \nu}=0$ in the background;
    \item $E^{TT\ \mu}_{\mu}=0$;
    \item $\nabla _{\mu }E^{TT\ \mu}_{\nu}=0$.
\end{itemize}
This has already been mentioned in section \ref{secpureAdS}.
If we want to consider a non-negligible contribution 
$E_{\mu \nu}^{corr}$ arising from $E^{TT}_{\mu
  \nu}$, this correction term $E_{\mu \nu}^{corr}$ has to satisfy the same
properties.
Since $E_{\mu \nu}^{corr}$ is transverse and traceless, we might think
of it as being derived from a conformally invariant action. Furthermore,
the correction $E_{\mu \nu}^{corr}$ should vanish for conformally
flat spacetimes, (since the effective theory with $E_{\mu \nu}^{TT}=0$
is exact for that case). We restrict ourselves to a correction
$E_{\mu \nu}^{corr}$, which is a functional of the metric only. Then the
most straightforward term derived from a conformally invariant
action, which vanishes for conformally flat spacetimes and is of
higher order in derivative, is:
\begin{align}
A_{\mu \nu } =&\frac{1}{2\sqrt{-g}}\frac{\delta }{\delta g^{\mu
\nu }}\int d^{4}x\sqrt{-g}\, C_{\alpha \beta \gamma \delta
}C^{\alpha \beta \gamma
  \delta }, \label{Auv} \\
A_{\mu \nu } =&
-\Box R_{\mu \nu}+\frac{1}{3}\nabla_{\mu}\nabla_{\nu}R
+\frac{1}{6}\Box R g_{\mu \nu} -\frac{1}{6}R^2 g_{\mu \nu}\\ 
&  +\frac{2}{3}R R_{\mu \nu}
+\frac{1}{2}R_{\alpha \beta}R^{\alpha \beta} g_{\mu \nu}
-2 R_{\mu \alpha \nu \beta}R^{\alpha \beta}. \notag
\end{align}
From the properties of the
four-dimensional Weyl tensor $C_{\alpha \beta \gamma \delta}$, $A_{\mu
  \nu }$
is indeed
traceless and vanishes for conformally flat spacetimes. 
There is therefore no reason why this term
would not be present in the electric part $E_{\mu \nu}$
 of the five-dimensional Weyl tensor. The purpose of this section
will be to study the effect when such a term is introduced:
\begin{align}
R_{\mu \nu }& =\frac{\kappa}{L} \left( T_{\mu \nu
  }-\frac{1}{2}Tg_{\mu \nu }\right)
-\frac{\kappa^2}{4}\left( T_{\mu }^{\alpha
}T_{\alpha \nu }-\frac{1}{3}TT_{\mu \nu }\right)
-E_{\mu \nu }, \label{modeinstein2} \\
E_{\mu \nu }& =\mathcal{E}_{\mu \nu }^{(L)}+E_{\mu \nu }^{TT}, \notag \\
E_{\mu \nu }^{TT}& =2 \alpha \, L^{2}A_{\mu \nu }+O(R_{\mu \nu
}^{3}),
\end{align}
where $\alpha $ is a dimensionless parameter that we assume to be
small. All through this study, we will work only up to first order in
$\alpha$. We may emphasize again that $A_{\mu \nu }$ does not alter
the background behaviour and will be present only at the
perturbative level.\bigskip

Further motivation for this study is given in appendix \ref{section KK}.
We show there that this correction correctly reproduces the
first KK mode for perturbations around two static branes
when matter is introduced only on one brane.
 Since this term is present in the two brane static limit, it seems
 natural to study its role within a brane inflation setup.
%
%
%
%
\subsection{Tensor perturbations}
We consider the previous scenario of brane inflation from section
\ref{Sectioninflation} and keep all terms in $L^2H^2$ so that the
analysis remains consistent at high energy. To start with, we consider
 tensor perturbations in detail.
We denote by $h_{ij}$ the three-dimensional tensor perturbation.
 By the same argument used previously, the
quadratic terms in the modified Einstein equation have only a
``background'' effect on the tensor perturbations. In particular,
they can be derived from the action:
\begin{align}
S^{(2)}_{t}=\int d^4x & \left( a^2\, \eta^{\mu \nu}\partial_{\mu}h^i_j\,
\partial_{\nu}h^j_i \right. \\
+& \left. \alpha\, L^2\, \eta^{\mu \nu}\eta^{\alpha
  \beta}\left(2\,\partial_{\mu \alpha}h^i_j\, \partial_{\nu
  \beta}h^j_i
-\partial_{\mu \nu}h^i_j\, \partial_{\alpha
  \beta}h^j_i
\right) \right). \notag
\end{align}
This gives rise to the equation of
motion for $h_{ij}$ (omitting the $ij$-indices for now):
\begin{eqnarray}
\ddot{h}+2\frac{\dot a}{a}\dot{h}-\nabla^2 h+\frac{\alpha
  L^2}{a^2}\left(\ddddot{h}-2\nabla^2 \ddot h+\nabla^4 h \right)=0 . \label{h''''}
\end{eqnarray}
We use here the assumption that the constant $\alpha$ is small and
expand $h$ as a series in $\alpha$: $h=h_0+\alpha\, h_1+\mathcal{O}(\alpha^2)$,
where $h_{0,1}$ do not depend on $\alpha$. To lowest order in
$\alpha$, we should recover the results from section \ref{section
  tensor1}. $h_0$ therefore satisfies:
$\ddot{h_0}+2\frac{\dot a}{a}\dot{h_0}-\nabla^2 h_0=0$. 
Keeping only the terms up to first
order in $\alpha$, the last term in eq. (\ref{h''''}) can be written as:
\begin{align}
\frac{\alpha L^2}{a^2} &\left(\ddddot{h}-2\nabla^2 \ddot h+\nabla^4 h \right)
=
\frac{\alpha L^2}{a^2} \left(\ddddot{h_0}-2\nabla^2 \ddot h_0+\nabla^4 h_0 \right)
+\mathcal{O}(\alpha^2)\\
&=\frac{\alpha L^2}{a^2} \Big(
-2\underbrace{\left[\frac{\dddot a}{a}-9\frac{\dot a \ddot a}{a^2}
+12 \frac{\dot a^3}{a^3} \right]}
_{
=2\dot a a H^2 \epsilon +\epsilon\, \mathcal{O}(\epsilon,\eta)
}
\dot h
+4\underbrace{\left[\frac{2\dot a^2}{a^2}-\frac{\ddot a}{a} \right]}
_{= a^2 H^2 \epsilon 
}
\nabla^2 h \Big)
+\mathcal{O}(\alpha^2).
\end{align}
Using this approximation, eq. (\ref{h''''}) simplifies considerably:
\begin{eqnarray}
\ddot{h}+
2\frac{\dot a}{a}\left(1-2\alpha L^2 H^2\, \epsilon \right)
\dot h-\left(1-4\alpha
  L^2 H^2\, \epsilon \right)\nabla^2 h=\mathcal{O}(\alpha^2). \label{h''2}
\end{eqnarray}
In this result,
the terms beyond first order in the slow-roll parameters have been omitted.
Eq. (\ref{h''2}) is consistent with the original eq. (\ref{h''''}) 
if we study only the terms up to first
order 
in $\alpha$ in the
expression of $h$.
This is an important assumption since the fourth order differential
equation reduces to a second order one, allowing us to specify only
two parameters on the initial Cauchy surface instead of four.
Using this assumption, the requirement that we recover a normalized
Minkowski vacuum when the modes are well inside the horizon is then
enough to specify $h_{ij}$. Similarly to the standard case, it is
simpler to study the associated variable $u$ instead:
\begin{align}
& z=a\left(1+\alpha L^2H^2\left(1+\epsilon \right)
\right), \label{zh2}\\
& u=zh,\\
& c^2_{h} =1-4 \epsilon \, \alpha L^2 H^2, \\
& \beta=2+3\epsilon \left(1-2\alpha L^2 H^2 \right),
\end{align}
so that the second order differential equation for the tensor
perturbations simplifies to:
\begin{eqnarray}
\ddot{u}+\left(c^2_{h}\, k^2-\frac{\beta}{\tau^2}\right)u=0,
\end{eqnarray}
up to first order in the slow-roll parameter $\epsilon$ and in
$\alpha$. \bigskip

The main point to notice is that the tensor modes do not
propagate at the speed of light anymore but at the speed of sound
$c_{h}$. The $C^2$-corrections (as introduced in $A_{\mu \nu}$) will
therefore modify the effective speed of linear perturbations on the
brane. There is a priori no reason for the constant $\alpha$ to be
positive. If $\alpha$ is negative, it will therefore be interesting to study whether or not
the speed of sound being effectively larger than the speed of light
from a four-dimensional point of view may result in
instabilities at higher order.

The amplitude of the tensor
perturbations at sound horizon crossing ($aH=c_{h} k$) is
multiplied by an extra mode-dependent factor:
\begin{eqnarray}
\mathcal{P}_g &\simeq&  \left. \frac{24}{M_4^4}
V\left(1+\frac{V}{2\mathcal{T}} \right)
\left(1- 2\alpha L^2H^2 \right)
 \right|_{\tau=\tau^*}. \label{tensorKK}
\end{eqnarray}
The effect of the $C^2$-corrections is extremely simple and
the extra term becomes negligible at long
 wavelengths as $k^2\tau^2 \ll 1$ and at low-energy, but it is not
 negligible at high energy.
Similarly, the tensor spectral index is modified by an extra
mode-dependent factor:
\begin{eqnarray}
n_T=-2\epsilon \left(1-2\alpha L^2H^2 \right).
\end{eqnarray}
Neglecting the slow-roll parameters, $H$ may be treated as constant
in this result. We can interpret the previous result as a background
redefinition:
\begin{eqnarray}
\bar{\epsilon}=\epsilon\, \left(1-2\alpha L^2H^2 \right),
\end{eqnarray}
even if the $C^2$-corrections did not perturb the background
behaviour.
%
%
\subsection{Scalar perturbations}
The scalar perturbations can be treated within the same philosophy as
the tensor perturbations so some of the details will be skipped.

In longitudinal gauge, for scalar isotropic perturbations,
%
\begin{eqnarray}
\begin{array}{l}
A_{\mu \nu }= \frac{2}{3\,a^{2}}\left[
\begin{array}{cc}
\nabla ^{4} & \nabla ^{2}\partial_i \\
\nabla ^{2}\partial_i & \
\left[
\frac 1 2 
\left( \partial_{\tau}^2-\nabla^2 \right)
\left(\partial_{ij}
-\nabla ^{2}\delta _{ij}\right)
+\partial_{ij}\partial_{\tau}^2
\right]
\end{array}
\right](\Psi+\Phi)
\end{array}
\end{eqnarray}
Introducing this term in the modified Einstein equation
(\ref{modeinstein2}),
the equation (\ref{equ}) for $u$ gets modified to:
\begin{align}
& u=\tilde{z}\Psi, \\
&  \ddot u+\left(c_{u}^2k^2-\frac{\tilde{\beta}}{\tau^2}
\right)u=0, \label{equuKK}
\end{align}
with the modified parameters:
\begin{align}
& \tilde{z}=z\, \left(1+\alpha \, z_1 \right),
\\
& z_1=\frac{4L^2H^2}{3}\left(1+\frac 1 2 \frac{L^2H^2}{1+L^2
H^2}\, \epsilon \right)
+2L^2H^2\left( 1+2\eta-\frac{1-\frac 1 2
L^2H^2}{1+L^2H^2}\, \epsilon
\right), \\
& c_{u}^2=
1+8\alpha L^2 H^2\left(1+\frac{2}{3}\frac{1+7/4 L^2 H^2}{1+L^2H^2}
\,  \epsilon\right), \\
& \tilde{\beta}=\beta+4\alpha L^2H^2\,  \epsilon.
\end{align}
In the next section, the exact result is developed in the low-energy
regime. In particular it is shown that the duality between slow-roll
and fast-roll conditions in the production of a scale invariant
spectrum remains valid when those kinds of corrections are taken into
account. For now, for simplicity, we consider the first order
terms in the slow-roll parameter only; in the previous result, terms
beyond first order in $\alpha$ or in the slow-roll
parameters have been omitted.\bigskip

We can notice here the same phenomenon as for the tensor
perturbations: the scalar perturbations do not propagate at the speed
of light any more but at the speed of sound $c_{u}$, which is not
the same speed as for tensor perturbations. \bigskip

The rest of the discussion remains the same. Assuming again that $\tilde
\beta$ may be treated as a constant, an analytical solution can be
found for (\ref{equuKK}) and the constants are chosen so that we
obtain Minkowski vacuum
$u_k \approx \frac{i\, e^{-i c_{u}k \tau}}{(2\, c_{u}k)^{3/2}}$ in
the regime where $c_{u}^2k^2 \tau^2 \gg \tilde{\beta}$. 
The expression between the scalar field perturbations $\delta \varphi$
an the variable $u$ is:
\begin{eqnarray}
\delta \varphi=-\frac{2L}{a\kappa}
\left[
\left(1+2\alpha L^2 H^2 +\mathcal{O}(\epsilon)
\right)\dot{u} +\frac{\ddot \varphi_0}{\dot \varphi
_0}\left(1+2\alpha L^2 (2\nabla^2+H^2 +\mathcal{O}(\epsilon)
)\right)u  
\right]
\end{eqnarray}
so that at short distances, the spacetime is locally flat and
 we recover the Bunch-Davis vacuum for the scalar field perturbations:
%
\begin{align}
\text{for } &  c_u^2k^2\tau^2 \gg \beta^2, \notag \\
 u_k & \approx \frac{i e^{-i \, c_{u}k
     \tau}}{(2\, c_{u} k)^{3/2}},\notag \\
\delta \varphi_k & \approx -\frac{L}{a \kappa }\frac{e^{-i c_{
       \varphi}k \tau}}{\sqrt{2 c_{ \varphi}k}}\left(1-\frac{i}
{c_{\varphi}k \tau} \right)\left(1+6 \alpha  L^2H^2 \right),\\
\text{with }\   c_u & =1+4 \alpha L^2 H^2 +\mathcal{O}(\epsilon)\ \
\text{ and }
\ \ c_{\varphi}=1+8 \alpha L^2 H^2+\mathcal{O}(\epsilon).\notag
\end{align}
We point out that the scalar field
perturbations
propagate at a
speed of sound slightly different from the speed of sound of the
Mukhanov scalar $u$.

Following
the same procedure as before, the spectral index for scalar
perturbations reads:
\begin{eqnarray}
n_S-1=-6\epsilon+2\eta-8\alpha L^2 H^2\, \epsilon,
\end{eqnarray}
which could be again interpreted as a redefinition of the slow roll-parameter:
\begin{eqnarray}
\tilde{\eta}=\eta +10 \alpha L^2 H^2 \,\epsilon.
\end{eqnarray}
The overall amplitude of the scalar
perturbations
gets a mode-dependent factor:
\begin{eqnarray}
\mathcal{P}
\sim \left. \frac{V^3\,
\left(1+\frac{V}{2\mathcal{T}}\right)^3} {M^6_4\, V_{,\varphi}^2} \,
   \left(1- 16\, \alpha \,
L^2 H^2
\right)
\right|_{\tau=\tau^{*}/c_{u}} . \label{Pu2}
\end{eqnarray}
We obtain the same kind of corrections as for tensor
perturbations (\ref{tensorKK}) which again might be significant at
high-energy. Since the scalar perturbation
propagates at a speed of sound $c_u$, there is
an extra factor when evaluating (\ref{Pu2}) due to the fact that the
modes exit the horizon at a different time:
$\left. V\,\left(1+\frac{V}{2\mathcal{T}}\right) \right|_{\tau^*/c_u}
\sim \left. V\,\left(1+\frac{V}{2\mathcal{T}}\right)\left(1-8\epsilon
\alpha L^2 H^2 \right) \right|_{\tau^*}$, but this difference
is negligible at long wavelengths.

 The ratio between
the amplitude of the tensor perturbations to the scalar ones 
acquires the
additional mode-dependent factor (to first order in $\epsilon$):
\begin{eqnarray}
r= \left. \frac{\epsilon}{1+V/\mathcal{T}}
 \left(1+16 \alpha
L^2 H^2
\right)\right|_{\tau \approx
 \tau^{*}}.
\end{eqnarray}
We may as well check that non-linear corrections remain small in the
context of this perturbative approach:
\begin{eqnarray}
\frac{\sqrt{\zeta^{(2)}_{\rho}}}{\zeta^{(1)}_{\rho}}\simeq
\left.
\frac{3}{\sqrt{2}}\,  \epsilon-8\sqrt 2 \alpha L^2
H^2 \right|_{\tau=\tau^*}.
\end{eqnarray}
We can see that the ratio is slightly modified by a mode-dependent
term which is {\it not} damped by the slow-roll parameter. Within
this approach the correction term is required to be small since we
are making an expansion in $\alpha$. However, a region can exist 
for which $\alpha$ might be small but still important compared to
$\epsilon$. In this case some deviations from non-gaussianity might
be observed. (We can point out that they will not be present
at low-energy, when the
terms $L^2H^2$ are negligible.) This is one of
the most interesting consequences of the addition of the
$C^2$-corrections within the context of brane inflation.
Another result is presented in the next section,
where we will
study the consequences of those corrections while considering the
production of perturbation with a fast-roll potential.
%
%
\subsection{Slow-Roll, Fast-Roll conditions
in the production of scale invariant spectrum}

It has been shown \cite{GKST,LPN} that within the low-energy
approximation, there is an exact duality between ``inflation''-like
and ``Ekpyrotic or Cyclic''-like potentials which give rise to the
same observational features (for scalar perturbations). We intend to
study here whether this duality is preserved when the
$C^2$-corrections are taken into account.
 In the setup of the previous section the scalar field was
confined on the brane. We will make the analogy between this case and
the case for which the scalar field can be
interpreted as the dilaton in a two brane Randall-Sundrum model.
In that case the dilaton does not live on a specific brane,
however from the four-dimensional effective point of view, the
formalism will be the same. In what follows we will work in a low-energy regime and
neglect the quadratic terms in the stress-energy tensor.
In the Cyclic scenario, the two boundary branes from the Randall-Sundrum model
are taken to be empty (at the time when the fluctuations
responsible for the structure are produced) and the effective
four-dimensional theory as derived in section \ref{BD}
is modified by the introduction of a
potential $V$:
\begin{eqnarray}
S=\int d^4 x\sqrt{-g} \left( \frac {L}{2 \kappa} R-\frac{1}{2}
\left(\partial \phi \right)^2 -V(\phi)\right). \label{S4dV}
\end{eqnarray}
$V$ represents an
interaction between the two branes which may come from bulk fields
such as in the Goldberger-Wise mechanism. The metrics on the branes
are conformally related to the effective one by the relations
(\ref{g+intermasofg4d}). The action
(\ref{S4dV}) gives some equations of motion which are
consistent with the background ones. However the addition of the
term $\alpha L^2 \, C_{\alpha \beta \gamma \delta} C^{ \alpha \beta
\gamma \delta}$ in this action will not alter the background
behaviour and thus will be a consistent term to consider. When
deriving the covariant four-dimensional effective action in section
\ref{BD}, there is indeed no reason why such a term could not have been
introduced. We will therefore consider the following action:
\begin{eqnarray}
S=\int d^4x  \sqrt{-g} \left( \frac{L}{2 \kappa}
R-\frac{1}{2}(\partial \phi)^2-V(\phi)+ \alpha L^2\,  C_{\alpha \beta \gamma
\delta} C^{\alpha \beta \gamma \delta} \right). \label{S4d+CC}
\end{eqnarray}
It is important to remember that the Einstein frame has no physical
significance here; it is in the sense of the theory translated back
into the brane frame that we think of those corrections. But
working in terms of (\ref{S4d+CC}) we may think of those
corrections in the same way as we have done so far and compare them
with the case of standard slow-roll inflation on the brane. \bigskip

Following the work of \cite{GKST}, in the low-energy limit, the
differential equation for the scalar modes may be expressed as:
\begin{eqnarray}
\ddot{u}+\left(\hat c_s^2 k^2-\frac{\hat{\beta}}{\tau^2} \right) u=0,
\label{equ3}
\end{eqnarray}
where $u$ is related to the curvature by
$u=\frac{a}{\dot{\varphi}_0} \left(1+\alpha\, \hat z \right)\Psi$,
with notation:
\begin{align}
& \epsilon=-\frac{\dot{H}}{a H^2}, \
\lambda^{(n)}=\frac{d^{n}\ln \epsilon}{ d \ln a\ ^n}, \\
& \hat{z}=\frac 4 3 \, \frac{L^2 k^2}{a^2}+
2L^2H^2 \left(1+3 \epsilon - \lambda^{(1)} \right),\\
& \hat c_s ^2=1+\frac{8}{3}\alpha L^2 H^2 \left(3+2 \epsilon \right),\\
& \hat \beta =-\frac{a^2H^2 \tau^2}{4}\left(
2\lambda^{(2)}-\left(2+ \lambda^{(1)}\right)\left(2\epsilon+ \lambda^{(1)}\right)
\right), \\
& +2\alpha L^2 a^2 \tau^2 H^4 \left(
\begin{array}{l}
\lambda^{(3)}-\lambda^{(2)}-\lambda^{(1)}\lambda^{(2)}-6 \lambda^{(2)}
\epsilon, \\
+2\epsilon+4 \epsilon^2-6\epsilon^3+3\epsilon\lambda^{(1)}
+11 \epsilon^2 \lambda^{(1)}
\end{array}
\right). \notag
\end{align}
Using the same arguments as in section \ref{Sectioninflation},
 eq. (\ref{equ3}) will admit a nearly scale-invariant solution if the variation of
 $\hat \beta$ is negligible and if $\hat \beta \ll 1$.
 If the parameter $\epsilon$ is considered constant as a first approximation
 (we set all the $\lambda^{(n)}$ to zero to start with),
 then by integration, $H=\frac{1}{(\epsilon -1)a \tau}$.
 The expression for $\hat \beta$ simplifies to:
\begin{align}
& \hat \beta=\frac{\epsilon}{(\epsilon-1)^2}\left[
1+4\alpha L^2 H^2 \left(1+2\epsilon-3 \epsilon^2 \right)  \right],\\
& H \sim \tau^{- \frac{\epsilon }{\epsilon -1}}.
 \end{align}
Setting the constant $\alpha$ to zero in this expression gives rise
to the standard result: $\hat \beta$ remains a small constant $\hat \beta
\ll 1$ for both $\epsilon \ll 1$ and $\epsilon \gg 1$. However,
adding the $C^2$-corrections in the effective theory gives rise to
the slightly more complicated situation: in the standard slow-roll
inflation case as studied in the previous section $\epsilon
\ll 1$, so that $\hat \beta \simeq \epsilon \left(1+4\alpha L^2 H^2
\right)$, with $H \simeq \text{constant}$ during inflation, giving
rise to the almost scale-invariant spectrum as discussed previously.\bigskip

For $\epsilon \gg 1$, on the other hand,
\begin{eqnarray}
&& \epsilon_{\text{\tiny{F}}}=\frac{1}{2 \epsilon} \ll 1, \\
&& \hat \beta=2\, \epsilon_{\text{\tiny{F}}}\, \left( 1-3\alpha L^2
H^2
 \epsilon_{\text{\tiny{F}}}
^{-2}\right).
\end{eqnarray}
$\epsilon_{\text{\tiny{F}}}$ denotes the fast-roll parameter as
introduced in \cite{GKST}. 

Two important problems arise here. First of all,
$\alpha L^2H^2$
 should be of the
same order of magnitude or smaller than $\epsilon_{\text{\tiny{F}}}^2$ to
avoid the correction term becoming large. If not, the procedure
breaks down as the correction becomes more important than the
``leading" term and higher order terms in $\alpha$ have to be
introduced as well as other corrections in the Weyl tensor.
In the low-energy limit the term $L^2H^2$ will
indeed be small, but not necessarily small compared to
$\epsilon_{\text{\tiny{F}}}$.
 The
second departure from the standard result arises from the
$\tau$-dependence of the correction term.
In the context of fast-roll, the Hubble parameter is not constant but varies as
$H\sim \tau^{-1}$.
 In general the corrections to $\hat \beta$ will not be
constant and it will not be possible
to follow
the usual derivation from eq.(\ref{equ}) to get the
expression (\ref{eqpsik}) and find a scale
invariant power spectrum. \bigskip

In the
Cyclic scenario, the large scale structure is produced while the
 Hubble parameter is still tiny and its variations small enough for the quantity
$L^2 H^2 \epsilon_{\text{\tiny{F}}} ^{-2}$ to remain
small and constant during the process. Those corrections will
therefore have a negligible effect on the power
spectrum of the perturbations.
It will be consistent to keep
treating $\hat \beta$ as a small constant, giving rise to the almost
scale-invariant spectrum:
\begin{eqnarray}
n_{S\text{, fast-roll}}-1=-4  \epsilon_{\text{\tiny{F}}}-4 \eta_{\text{\tiny{F}}}
+12\left. \alpha L^{2} H^2 \epsilon_{\text{\tiny{F}}}^{-1}\right|_{\tau ^*},
\end{eqnarray}
with the second fast-roll parameter as defined in \cite{GKST}:
$\eta_{\text{\tiny{F}}}=1-\eta \, \epsilon_{\text{\tiny{F}}}$.\bigskip

However in a more general case of braneworld cosmology,
if we consider a general potential
satisfying fast-roll conditions, we will generically expect to see
some departure from a nearly-scale invariant spectrum when some
$C^2$-corrections are introduced in the Weyl tensor.


\section{Conclusion \label{sectionconclusion}}

In the first part of this work we pointed out the presence of a
transverse part of the Weyl tensor, that vanishes for the background
(but does not necessarily cancel in general).
We began with the assumption
 that this part of the Weyl tensor could
 be neglected. This was shown to be a valid approximation
 in the case of adiabatic
 scalar perturbations.\bigskip

 Using this
 assumption we solved the modified Einstein equation for
 brane inflation driven by a scalar field in a slow-roll
 potential.
The corrections coming from the quadratic terms in the stress-energy
were tracked throughout.
 We showed that the perturbations are
 anisotropic in longitudinal gauge, in contrast to the case of standard
 inflation.
We showed how to extend the standard inflation variables and
 parameters to accommodate the behaviour of the
 quadratic term. The corrections to these terms arise from a purely
 background effect. Indeed, assuming adiabaticity, the corrections can
 only
 influence the background. For a given inflation potential, the power
 spectrum of both the tensor and the scalar perturbations are redder
 than in the normal four-dimensional case. Compared to scalar
 perturbations, the tensor perturbation amplitude is weaker.
However for scalar perturbations, this model can
be reinterpreted as
  standard four-dimensional inflation with a redefined potential,
 giving rise to the same astrophysical observations.
  These results have already been well-understood in a
 number of papers
 \cite{chaoticinfationbrane,PotLiddle01,Calcagni:2003sg,
 Calcagni:2004bh,maartens,LMSW,padi}.
 However our study is an
 important consistency check as it enables us to verify our
 prescription and to extend it.
 The relations between the brane inflation variables and the redefined
 standard inflation variables have been
 given with precision, and our
 results are reliable up to second order in the slow-roll parameters.
Another important feature of our four-dimensional effective theory is
 its straightforward extension to more interesting and realistic
 scenarios where
 both boundary branes have their own dynamics. This opens the
 possibility of studying a
 large range of braneworld scenarios. \bigskip

To understand the contribution of the Weyl tensor,
we have extended what we knew of its background
behaviour to knowledge of its behaviour in a
quasi-static limit when the extra-dimension is finite.
 In that limit, when matter is introduced on
one brane, the first Kaluza-Klein mode
can be modelled by a tensor $A_{\mu \nu}$
which can be expressed in terms of local four-dimensional
quantities. Motivated by this result we have studied the contribution
of $A_{\mu \nu}$ in two particular examples.\bigskip

First we analyzed the contribution of the tensor $A_{\mu \nu}$ in a
model of brane inflation where again the typical energy scale was
important compared to the brane tension. Several new features were
 observed which, as far as we are aware,
 extend previous results in chaotic brane
inflation.
First of all, both tensor and scalar
modes propagate at a speed different from the speed of
light. The corrections bring a new mode-dependent term
 in the amplitude of both tensor and
scalar perturbations which does not compensate in their ratio and
which could be important at high energy.
There is as well a new mode-dependent
term in the estimation of the
non-gaussianity, proportional to the small constant $\alpha$
we have introduced. This term is not damped by a slow-roll parameter. This is a
critical
 feature,
addressing the issue of the importance of non-gaussianity
when those corrections are introduced.
Our prescription only makes sense when
the constant $\alpha$ is small, so
that we can not address the issue of non-gaussianity outside this regime,
 however there could be some situations for which the new term
 in the non-gaussianity estimation might be the leading one. In that case,
the cubic terms might not be small compared to the
quadratic ones, and a perturbative approach might not be sensible.
These are important features that have to be considered seriously in
order to distinguish between purely four-dimensional
inflation and brane inflation. Comparing these results with
observations might give a constraint
on the order of magnitude of the constant $\alpha$. \bigskip

In the second example, we used the formalism to study how the duality
relating density perturbations in expanding and contracting Friedmann
cosmologies was affected by the introduction of the tensor $A_{\mu
  \nu}$.
 We compared the production of a scale-invariant spectrum in a model 
of ``slow-roll''
inflation where the typical energy scale was much smaller than the
brane tension, with an Ekpyrotic or Cyclic model for which
the scalar field was evolving in a potential
satisfying ``fast-roll'' conditions. The first order corrections,
proportional
to the constant $\alpha$,
 become negligible in the ``slow-roll'' limit but could be large in
a general ``fast-roll'' limit.
    We therefore recover the nearly
scale-invariant spectrum in the ``slow-roll'' inflation but the
situation becomes more complicated in a general ``fast-roll''
scenario. In this case, unless some assumptions are
made
during the production
of perturbations responsible for the observed structure
(as in the case of Cyclic-model potentials), the
departure from a scale-invariant spectrum could become more important
when these corrections interfere.
This is an important new result enabling us to
differentiate between the ``fast-roll'' and ``slow-roll'' scenarios.
However, when comparing the specific case of the Cyclic
model with low-energy inflation,
 the corrections are negligible.  \bigskip


The main limitation of this study is the assumption made for the behaviour
of the
homogeneous part of the Weyl tensor.
At each step, its behaviour has been imposed by hand.
We have tried however to go beyond the usual assumptions and
incorporate some effects of
the brane nature of our theory by analogy with what is already known from
purely four-dimensional theories. To be
completely rigorous, one should ultimately try to
attack the five-dimensional problem directly.
Nonetheless, we hope that our formalism can be used to give greater analytical
insight into typical braneworld effects. 
\section*{\normalsize{Acknowledgements}}
\vskip -2pt
I am grateful to my supervisor N.~Turok.
I wish to thank as well
A.~Davis, C.~van de Bruck, P.~Steinhardt,
S.~Gratton, P.~McFadden,
T.~Wiseman and A.~Tolley
for useful discussions.
This work was supported by Girton College, Cambridge, together with
the support of ORS and COT awards from Cambridge University.\bigskip

%
%
\appendix
\setcounter{equation}{0}
\renewcommand{\theequation}{\thesection.\arabic{equation}}
\section{Covariant formalism including the $\rho ^{2}$
 terms  \label{section p2 terms}}
Our starting point will be the assumption that for long wavelength
adiabatic perturbations, the metrics on each brane remain conformal related
as in (\ref{g+intermasofg4d}),
\begin{eqnarray}
g_{\mu \nu }^{(-)}=\Psi ^{2}g_{\mu \nu }^{(+)}, \label{conformal
assumption}
\end{eqnarray}
where the conformal factor $\Psi $ may be expressed in terms of
the minimally coupled scalar field $\phi$ of (\ref{4daction})
 as $\Psi =-\tanh(\phi
/\sqrt{6})$. This assumption is verified for cosmological
metrics and as a first step we will assume it will remain valid
for long wavelength adiabatic perturbations.

Using this assumption, the Ricci tensor on the negative tension
brane can be expressed in terms of the Ricci tensor on the positive
one:
\begin{eqnarray}
\Psi ^{2}R_{\mu \nu }^{(-)}=\Psi ^{2}R_{\mu \nu }^{(+)}-2\Psi \nabla
_{\mu }\nabla _{\nu }\Psi +4\;\partial _{\mu }\Psi \;\partial _{\nu
}\Psi -g_{\mu \nu }^{(+)}\left( \Psi \Box \Psi +\left| \nabla \Psi
\right| ^{2}\right), \label{R- as R+}
\end{eqnarray}
where all derivatives and contractions are taken with respect to the
metric on the positive tension brane. This will be our convention
throughout this section unless otherwise specified. Taking the
trace of eq. (\ref{R- as R+}), we get the equation of motion of the
scalar field $\Psi $:
\begin{eqnarray}
\Box \Psi =\frac{1}{6}\left( R^{(+)}-\Psi ^{2}R^{(-)}\right) \Psi,
\label{eq motion of psi}
\end{eqnarray}
where $R^{(-)}=g^{(-)\mu \nu }R_{\mu \nu }^{(-)}$.\ We notice that
if the negative (resp. positive) tension brane is empty $R^{(-)}=0$,
the scalar field $\Psi $ (resp. $1/ \Psi$) is conformally invariant
with respect to the positive (resp. negative) tension brane metric.\bigskip

As already mentioned in section \ref{secpureAdS}, the projected Ricci
tensor on each brane can be expressed as:
\begin{eqnarray}
\begin{array}{l}
R_{\mu \nu }^{(\pm )}=\pm \frac{\kappa}{L} \left( T_{\mu \nu
}^{(\pm )}-\frac{1}{2}T^{(\pm )}g_{\mu \nu }^{(\pm )}\right)
-\frac{\kappa ^2}{4}\left( T_{\mu }^{(\pm )\alpha }T_{\alpha \nu
}^{(\pm )}-\frac{1}{3}T^{(\pm )}T_{\mu \nu }^{(\pm )}\right) -E_{\mu
\nu }^{(\pm )}
\end{array}  \label{Ricci2}
\end{eqnarray}
with $E_{\mu \nu }^{(\pm )}$ as in (\ref{Euv+-}). For the
background, $E_{\mu \nu }^{(-)}$ and $E_{\mu \nu }^{(+)}$\ are
therefore related:
\begin{eqnarray}
\Psi ^{2}E_{\mu \nu }^{(-)}=E_{\mu \nu }^{(+)}.  \label{deltaE=0}
\end{eqnarray}
A priori, $E_{\mu \nu }^{(\pm )}$ are not determined except for the
following properties:
\begin{eqnarray}
g^{(\pm )\mu \nu }E_{\mu \nu }^{(\pm )}=0,  \label{Etraceless}
\end{eqnarray}
and the fact that their divergences must be consistent with the
Bianchi identities on each brane $\nabla ^{(\pm)}_{\mu} G^{(\pm) \,
\mu}_{\nu}=0$.

Using eq. (\ref{Ricci2}), we can construct the quantity $\left(
R_{\mu \nu }^{(+)}-\Psi ^{2}R_{\mu \nu }^{(-)}\right) $:
\begin{eqnarray}
\left( R_{\mu \nu }^{(+)}-\Psi ^{2}R_{\mu \nu }^{(-)}\right) =\left(
\Pi _{\mu \nu }^{(+)}-\Psi ^{2}\Pi _{\mu \nu }^{(-)}\right) -\Delta
E_{\mu \nu }, \label{diffR}
\end{eqnarray}
where
\begin{eqnarray}
\Delta E_{\mu \nu }=\left( E_{\mu \nu }^{(+)}-\Psi
^{2}E_{\mu  \nu }^{(-)}\right),
\end{eqnarray}
 and $ \Pi _{\mu \nu }^{(\pm )}=\pm
\frac{\kappa}{L}\left( T_{\mu \nu }^{(\pm )}-\frac{1}{2} T^{(\pm
)}g_{\mu \nu }^{(\pm )}\right) -\frac{\kappa ^2}{4}\left( T_{\mu
}^{(\pm )\alpha }T_{\alpha \nu }^{(\pm )}-\frac{1}{3}T^{(\pm
)}T_{\mu \nu }^{(\pm )}\right) $.
In what follows we will write
$\tilde{\Pi}_{\quad \nu }^{(\pm )\mu }=\Pi _{\quad \nu }^{(\pm )\mu
}- \frac{1}{2}\Pi ^{(\pm )}\delta _{\nu }^{\mu }$ with $\Pi _{\quad
\nu }^{(\pm) \mu }\equiv g^{(\pm)\mu \alpha \,}\Pi _{\alpha \nu
}^{(\pm)}$. \bigskip

Using the relation (\ref{R- as R+}) between $R_{\mu \nu }^{(-)}$ and
$R_{\mu \nu }^{(+)}$, the Ricci tensor on the positive tension brane
may be expressed in terms of the scalar field $\Psi $:
\begin{eqnarray}
R_{\mu \nu }^{(+)}=\Pi _{\mu \nu }^{(+)}+U_{\mu \nu }^{\Psi} -\Delta
E_{\mu \nu },  \label{cov Ricci+}
\end{eqnarray}
where $U_{\mu \nu }^{\Psi}=-2\Psi \nabla _{\mu }\nabla _{\nu }\Psi
+4\;\partial _{\mu }\Psi \;\partial _{\nu }\Psi -g_{\mu \nu
}^{(+)}\left( \Psi \Box \Psi +\left| \nabla \Psi \right| ^{2}\right)
+\Psi ^{2}R_{\mu \nu }^{(+)}-\Psi ^{2}\Pi _{\mu \nu }^{(-)}$
satisfies:
\begin{align}
g^{(+)\mu \nu \,}U_{\mu \nu }^{\Psi} = & 0, \\
\nabla _{\mu }^{(+)}U_{\quad \nu }^{\Psi \, \mu } =&-\Psi
^{2}\nabla _{\mu }^{(-)}\tilde{\Pi}_{\quad \nu }^{(-)\mu }.
\end{align}
The covariant derivatives are taken with respect to $g_{\mu \nu
}^{(+)}$, unless otherwise stated.

Since the brane metrics are conformal to each other, the traceless
property of $E_{\mu \nu }^{(\pm )}$ can be extended to $\Delta E_{\mu
\nu }$. Furthermore, from the Bianchi identity the divergence of
$\Delta E_{\mu \nu }$ is:
\begin{eqnarray}
\nabla _{\mu }^{(+)}\left( g^{(+)\mu \alpha }\Delta E_{\alpha \nu
}\right) =\nabla _{\mu }^{(+)}\tilde{\Pi}_{\quad \nu }^{(+)\mu
}-\Psi ^{2}\nabla _{\mu }^{(-)}\tilde{\Pi}_{\quad \nu }^{(-)\mu }.
\label{divofE}
\end{eqnarray}
The traceless tensor $\Delta E_{\mu \nu }$ can be decomposed into a
vector part and a traceless, divergenceless tensor:
\begin{align}
\Delta E_{\mu \nu } &=\mathcal{E}_{\mu \nu }+E_{\mu \nu }^{TT}, \\
\mathcal{E}_{\mu \nu } &=\nabla _{\mu } \mathcal{E}_{\nu }+\nabla
_{\nu } \mathcal{E}_{\mu }-\frac{1}{2}g_{\mu \nu }^{(+)}\nabla
_{\alpha }\mathcal{E}
^{\alpha }, \\
\nabla _{\mu }E_{\quad \nu }^{TT\;\mu }&=0 \text{ and } E_{\quad
\mu}^{TT
  \;\mu }=0.
\end{align}
The vector $\mathcal{E}_{\mu } = \mathcal{E}^{(+)}_{\mu } -
\mathcal{E}^{(-)}_{\mu }$ \ may be determined using the relation for
the divergence of $\Delta E_{\nu }^{\mu }$ in eq.(\ref{divofE}):
\begin{align}
\mathcal{E}_{\mu \nu}&=\mathcal{E}^{(+)}_{\mu
\nu}-\mathcal{E}^{(-)}_{\mu
  \nu}, \\
\nabla ^{(+)}_{\mu } \mathcal{E}^{(+)\mu}_{\nu}&=\nabla _{\mu
}^{(+)}\tilde{\Pi}
_{\quad \nu }^{(+)\mu },\\
\nabla ^{(+)}_{\mu } \mathcal{E}^{(-)\mu}_{\nu}&=\Psi ^{2}\nabla
_{\mu }^{(-)}\tilde{\Pi}_{\quad \nu }^{(-)\mu }.
\end{align}
We consider that all the divergenceless part of $ \Delta E_{\mu \nu
}$ is contained in $ E_{\mu \nu }^{TT}$. Namely, $\nabla _{\mu
}\mathcal{E}_{\nu }^{(\pm) \mu }=0 \, \Leftrightarrow \,
\mathcal{E}^{\pm}_{\mu }=0$. Adding the second term to the tensor
$U^{\Psi}_{\mu \nu}$, we obtain the conserved and traceless tensor:
\begin{eqnarray}
&& T^{\Psi \text{ eff}}_{\mu \nu}=U^{\Psi}_{\mu
  \nu}+\mathcal{E}^{(-)}_{\mu \nu}, \\
&& T_{\mu \nu }^{\Psi \text{ eff}}=
4\partial _{\mu }\Psi \;\partial _{\nu }\Psi -2\Psi \nabla
_{\mu }\nabla _{\nu }\Psi +g_{\mu \nu }^{(+)}\left( 2\Psi \Box \Psi
-\left| \nabla \Psi \right| ^{2}\right) \label{Tuv eff} \\
&& \phantom{T_{\mu \nu }^{\Psi \text{ eff}}=}
+G_{\mu \nu }^{(+)}\Psi^2-\Psi
^{2}\tilde{\Pi}_{\mu \nu }^{(-)}+\mathcal{E}^{(-)}_{\mu \nu}, \notag \\
&& g^{(+)\mu \nu} T^{\Psi \text{ eff}}_{\mu \nu } = 0,
\label{Tuv traceless} \\
&& \nabla _{\mu }^{(+)} T^{\Psi \text{ eff}\, \mu}_{\quad \nu } = 0.
\label{Tuv conserved}
\end{eqnarray}
It is interesting to note that when the negative tension brane is
empty of matter and radiation, $\Psi ^{2}\tilde{\Pi}_{\mu \nu
}^{(-)}=\mathcal{E}^{(-)}_{\mu \nu}=0 $ and $T^{\Psi \text{
eff}}_{\mu
  \nu}$ is precisely the stress energy tensor of the scalar field $\Psi$
which would then be conformally invariant by (\ref{eq motion of
psi}). When matter is present on the negative tension brane, the
traceless and divergenceless properties of the pseudo-stress-energy
tensor are still satisfied. Namely some $T^{(-)}_{\mu \nu}$
contributions must be  added to the stress-energy tensor of $\Psi$ to
accommodate the fact that $\Psi$ is no longer conformally invariant
with respect to the
positive tension brane as mentioned in (\ref{eq motion of psi}).

The only part in (\ref{cov Ricci+}) that remains undetermined is
therefore the traceless and divergenceless $E_{\mu \nu }^{TT}$
contribution in $\Delta E_{\mu \nu }$. From the relation
(\ref{deltaE=0}), $\Delta E_{\mu \nu }=E^{(+)}_{\mu
\nu}-\Psi^2E^{(-)}_{\mu \nu}=0$ for cosmological metrics. So
$E_{\mu \nu }^{TT}$ vanishes in the background.
Following the argument of section \ref{secpureAdS}, for the purpose
of long wavelength adiabatic perturbations, it is
therefore consistent to set $E_{\mu \nu }^{TT}$ to zero.

Thanks to the Gauss-Codacci formalism and to the conformal
assumption (\ref{conformal assumption}), it is therefore possible to
find an expression for the Weyl tensor $E_{\mu \nu}$ up to a
conserved traceless tensor that shall be neglected for our regime of
interest. Comparing equation (\ref{Ricci2}) with (\ref{cov Ricci+}),
the expression of the Weyl tensor is:
\begin{eqnarray}
E^{(+)}_{\mu \nu}= -T^{\Psi \text{ eff}}_{\mu
\nu}+\mathcal{E}^{(+)}_{\mu \nu}. \label{Euv cov}
\end{eqnarray}
In this regime, the metric on both branes can be found by solving the
set of equations:
\begin{eqnarray}
\begin{array}{l}
%
R_{\mu \nu }^{(+)}=\left( \Pi _{\mu \nu }^{(+)}
-\mathcal{E}^{(+)}_{\mu \nu }\right)+T_{\mu \nu }^{\Psi
\text{ eff}}, \smallskip \\
%
\Box \Psi =\frac{1}{6}\left( R^{(+)}-\Psi ^{2}\Pi ^{(-)}\right) \Psi,
\smallskip \\
\end{array} \label{CloseSystem}
\end{eqnarray}
with
\begin{eqnarray}
\left\{
\begin{array}{l}
%
%
g_{\mu \nu }^{(-)}=\Psi ^{2}g_{\mu \nu }^{(+)} \smallskip \\
%
%
\Pi _{\mu \nu }^{(\pm )}=\pm \frac{\kappa}{L} \left( T_{\mu \nu
}^{(\pm )}-\frac{1}{2} T^{(\pm )}g_{\mu \nu }^{(\pm )}\right)
-\frac{\kappa ^2}{4}\left( T_{\mu }^{(\pm )\alpha }T_{\alpha \nu
}^{(\pm )}-\frac{1}{3}T^{(\pm )}T_{\mu \nu }^{(\pm
)}\right) \smallskip \\
%
T_{\mu \nu }^{\Psi \text{ eff}}=
4\partial _{\mu }\Psi \;\partial _{\nu }\Psi -2\Psi \nabla
_{\mu }\nabla _{\nu }\Psi +g_{\mu \nu }^{(+)}\left( 2\Psi \Box \Psi
-\left| \nabla \Psi \right| ^{2}\right) 
\\
\phantom{T_{\mu \nu }^{\Psi \text{ eff}}=}
+\Psi^{2}G_{\mu \nu }^{(+)}
-\Psi
^{2}\tilde{\Pi}_{\mu \nu }^{(-)}+\mathcal{E}^{(-)}_{\mu \nu}
\smallskip  \\
%
\mathcal{E}^{(\pm)}_{\mu \nu } =\nabla _{\mu }
\mathcal{E}^{(\pm)}_{\nu }+\nabla _{\nu } \mathcal{E}_{\mu
}^{(\pm)}-\frac{1}{2}g_{\mu \nu } \nabla _{\alpha }\mathcal{E}
^{(\pm) \alpha } \smallskip \\
%
\nabla _{\mu }\mathcal{E}_{\nu }^{(+)\mu }=\nabla _{\mu
}^{(+)}\tilde{\Pi}
_{\quad \nu }^{(+)\mu } \smallskip \\
%
\nabla _{\mu }\mathcal{E}_{\nu }^{(-) \mu }=\Psi ^{2}\nabla _{\mu
}^{(-)}\tilde{\Pi}_{\quad \nu }^{(-)\mu }  \smallskip
\end{array} \right. \label{def}
\end{eqnarray}
with $\Pi^{(\pm)}=g^{(\pm)\mu \nu} \Pi^{(\pm)}_{\mu \nu}$,
$\tilde{\Pi}_{\quad \nu }^{(\pm )\mu }=\Pi _{\quad \nu }^{(\pm )\mu
}- \frac{1}{2}\Pi ^{(\pm )}\delta _{\nu }^{\mu }$ and any other
contraction and derivatives done with respect to $g_{\mu \nu
}^{(+)}$. \bigskip

These may look formidable, we have an equation
of motion for a non-minimally coupled scalar field $\Psi$ and
equations for the metric $g_{\mu \nu}^{(+)}$ sourced in a highly 
non-trivial manner by $\Psi$ and the stress-energy on each brane.
However, we argue that the system of equations (\ref{CloseSystem}) with
(\ref{def}) is
self-consistent and allows us to calculate the long-wavelength
adiabatic perturbations when the $\rho ^{2}$ terms play a
significant role and cannot be neglected. It is perfectly consistent
with previous work done in the low-energy limit. Indeed, in the
limit where $\rho_{\pm} \ll \mathcal{T}$, the usual moduli space results are
obtained. \bigskip

The extension of this work to one brane in a non-compact
extra-dimension as described in the Randall-Sundrum II model is
straightforward by taking the limit $\Psi \rightarrow 0$.  \bigskip

This work relies on the important assumption that the branes remain
conformal to each other. As we shall see in the following, this 
assumption is
not true when Kaluza-Klein corrections are taken into account;
however we may deal with this fact by introducing a different
projected Weyl tensor on each brane.

\setcounter{equation}{0}

\section{First Kaluza-Klein Mode in the Effective Theory for Static
  Branes \label{section KK}}

\subsection{KK Corrections on the RS Model around Static
  Branes \label{sectionGarrigaTanaka}}
We consider two static flat branes of positive and negative
tension, embedded
in a five-dimensional Anti-de-Sitter (AdS) space with cosmological
constant $\Lambda=-\frac{6}{\kappa L^2}$.
 The bulk metric is:
\begin{eqnarray}
ds^2=dy^2+e ^{-2 |y|/L} \eta _{\mu \nu} dx^{\mu}dx^{\nu},
\end{eqnarray}
with the branes located at $y\equiv d^{\pm}$.
Both branes are subject to a $Z_2$ -reflection symmetry.
For this background solution, the induced metric on each brane
is flat
$\gamma_{\mu \nu}^{(\pm)}=e^{-2\frac{d^{\pm}}{L}} \eta_{\mu \nu}$.
We shall denote by $\Psi_0=e^{-d/L}$
the constant conformal factor relating the metrics on the two branes
$\gamma_{\mu \nu}^{(-)}=\Psi_{0}^2\, \gamma_{\mu \nu}^{(+)}$.\bigskip

Following the procedure of
\cite{GT}, we consider the metric perturbations sourced by
matter confined on the positive tension brane, with stress-energy tensor
$T_{\mu \nu}$.
 Since matter is introduced as a
 perturbation, to first order, the matter fields confined on
 the brane are living on the background flat metric $\gamma^{(+)}_{\mu
 \nu}$.

As considered in \cite{GT}, we shall first solve the equations of motion for the
perturbed metric in Randall-Sundrum (RS) gauge. In this gauge, the position
of the positive tension brane is not fixed.
We denote by $\delta y$ the deviation from its fixed background
 location. In order to express the perturbed metric induced on
each brane,
we will change to Gaussian
normal coordinates.

 When the distance between the two
branes is finite, the perturbed metric on the branes can be
expanded in momentum space. We shall be interested in the first order
correction to the zero mode. \bigskip

In RS gauge, the perturbed metric is:
\begin{eqnarray}
ds^2=dy^2+\left(e ^{-2 |y|/L} \eta _{\mu \nu}+h_{\mu \nu}
\right)dx^{\mu}dx^{\nu} \\
\text{with } h_{\mu}^{\mu}=0 \text{ and } h^{\mu}_{\nu , \mu}=0,
\label{RSgauge}
\end{eqnarray}
where $h_{\mu \nu}$ is transverse and traceless with respect to the
background metric $\gamma_{\mu \nu}=e ^{-2 |y|/L} \eta _{\mu \nu}$.
In this gauge, the five-dimensional Einstein equation and the Isra\"{e}l
matching conditions read:
\begin{eqnarray}
&& \left[e ^{2 y/L} \Box +\partial ^2
  _ y-\frac{4}{L^2} \right]h_{\mu \nu}=0 \ \text{ for }d^+<y<d^- , \label{diffeq}\\
&& \left[\partial_{y}+\frac{2}{L} \right]h_{\mu \nu}=\left\{\begin{array}{cl}
-\kappa \ \Sigma _{\mu \nu}& \text{ at }y=d^+ \\
0 & \text{ at }y=d^-
\end{array} \right., \label{bdycond}
\end{eqnarray}
where $\Box$ is the Laplacian on the four-dimensional Minkowski
space. The tensor
 $\Sigma _{\mu \nu}$ is a functional
 of the matter on the positive brane:
$\Sigma _{\mu \nu}=T_{\mu
  \nu}-\frac{1}{3}T \gamma^{(+)} _{\mu \nu}
-\frac{2}{\kappa}\,  \delta y_{,\, \mu \nu}$,
with $T=\gamma^{(+)\, \mu \nu}\, T_{\mu \nu}$.
From eqs. (\ref{bdycond}) and (\ref{RSgauge}), the tensor
$\Sigma _{\mu \nu}$
must be transverse and traceless with respect
  to its background flat metric $\gamma_{\mu \nu}^{(+)}$.  The
  deviation $\delta y$ from the background position is
  therefore
given by $\delta
y=-\frac{\kappa}{6}\frac{1}{\Box^{(+)}}T$ :
\begin{eqnarray}
\Sigma_{\mu \nu}=T_{\mu
  \nu}-\frac{1}{3}T \gamma^{(+)} _{\mu \nu}
+\frac{1}{3\, \Box ^{(+)}}T_{,\, \mu \nu}.
\label{defSigma}
\end{eqnarray}
 Since the background metric is flat,
$\Box^{(+)}=
e^{2\, \frac{d^+}{L}}\Box$. \bigskip

The differential equation (\ref{diffeq}) with boundary conditions
(\ref{bdycond})
can be solved, giving the expression for
 the perturbed metric in RS gauge:
\begin{eqnarray}
h_{\mu \nu}(y,x^{\mu})
=\kappa\,  \hat F(y) \Sigma _{\mu \nu}
. \label{huv}
\end{eqnarray}
In what follows, for simplicity, we take $d^{+}=0$ and $d^{-}=d$. In
that case,
the operator $\hat F$ can be expressed as:
\begin{eqnarray}
\begin{array}{l}
\hat F (y)=
\frac{1}{\sqrt{-\Box}} 
\left[
\frac{I_2(e^{y/L}L\sqrt{-\Box})
K_1(e^{d/L}L\sqrt{-\Box})+
I_1(e^{d/L}L\sqrt{-\Box})K_2(e^{y/L}L\sqrt{-\Box})}
{I_1(e^{d/L}L\sqrt{-\Box})K_1(L\sqrt{-\Box})-
I_1(L\sqrt{-\Box})K_1(e^{d/L}L\sqrt{-\Box})}
\right],
\end{array}
\label{OperatorF}
\end{eqnarray}
with $I_n$ (resp. $K_n$) the $n^{\text{th}}$ Bessel function of first
(resp. second) kind. We may notice that when $d^+=0$, $\gamma_{\mu
  \nu}^{(+)}=\eta_{\mu \nu}$ and
$\Box^{(+)}=\Box$.\smallskip

Since in RS gauge, the negative tension brane remains located at
$y\equiv d$, $h_{\mu \nu}(x^\mu,y=d)$ is the metric perturbation
induced on that brane. However this is not the case for the positive
tension brane.
In order to find the induced metric on that
  brane, we need to
perform a gauge transformation and work in terms of the Gaussian
  normal (GN$^{(+)}$) gauge for this brane:
\begin{eqnarray}
\left\{ \begin{array}{l}
 \bar{y}=y-\delta y \\
 \bar{x}^{\mu}=x^{\mu}
+\zeta^{ \mu}(x^\nu).
\end{array} \right. \label{gaugetrans}
\end{eqnarray}
When working in the GN$^{(+)}$ gauge, with coordinates
$(\bar{y},\bar{x}^{\mu})$, the positive tension brane is
located at $\bar{y} \equiv 0$.
Performing the gauge transformation (\ref{gaugetrans}) with
$\delta
y=-\frac{\kappa}{6}\frac{1}{\Box}T$,
the perturbed
metric induced on the positive tension brane is given by:
\begin{eqnarray}
\bar{h}^{+}_{\mu \nu}(\bar{y}\equiv 0)
=h_{\mu \nu}(y=0)
+\frac{\kappa}{3L}\eta_{\mu \nu}
\frac{1}{\Box}T-\zeta_{(\mu , \nu)},
\end{eqnarray}
where the indices are raised and lowered
with the metric $\eta^{\mu \nu}$.
We may fix the remaining degrees of freedom
by imposing the gauge choice:
\begin{eqnarray}
\bar{h}^{\mu} _{\,  \nu ,\mu}
=\frac{1}{2} \bar{h}^{\mu} _{\, \mu , \nu}
 \text{ at }\bar{y}=0.
\end{eqnarray}
 This gauge choice corresponds to the de Donder gauge and is obtained
 by fixing $\zeta^{ \mu}(x^{\nu})$ such that
$\zeta_{\mu}=-\frac{\kappa}{3L \Box^{2}} T_{,\, \mu}$.\\
In de Donder gauge, the metric perturbation induced on the positive
tension brane is therefore:
\begin{eqnarray}
\bar{h}^{(+)}_{\mu \nu}=h_{\mu \nu}(y=0)
+\frac{\kappa}{3L}\eta_{\mu
  \nu}\frac{1}{\Box}T
+\frac{2\kappa}{3 L}\frac{1}{\Box^{\,2}} T_{, \mu \nu}.
\end{eqnarray}
The perturbed metric on both branes is therefore:
\begin{eqnarray}
&& \bar{h}^{(+)}_{\mu \nu}=\kappa \left( \hat F(0) \Sigma _{\mu \nu}
+\frac{1}{3L}\eta_{\mu
  \nu}\frac{1}{\Box}T
+\frac{2}{3 L}\frac{1}{\Box^2} T_{, \mu \nu}
\right),\\
&& \bar{h}^{(-)}_{\mu \nu}=\kappa\, \hat F(d) \Sigma_{\mu \nu}.
\end{eqnarray}
In what follows we shall be interested in the first order expansion in
derivatives (in $ L^2 \Box $),
 of the operator $\hat F$.
The expression (\ref{OperatorF}) of $\hat F$ has a derivative
expansion:
\begin{eqnarray}
& \hat F(y)=\frac{2L\, e^{-2y/L}}{1-e^{-2d/L}}
\left[
-\frac{1}{L^2\Box}+ f(y)+\mathcal{O}(L^2 \Box)
\right],\label{expF}\\
& f(y)=\frac{d/2L}{1-e^{-2d/L}}-\frac{1}{8}
\left(1-e^{2(2y -d)/L}+2e^{2y/L}
\right).\label{deff}
\end{eqnarray}
The first term in (\ref{expF}) can be interpreted as the zero mode and
the second one as the first Kaluza-Klein (KK) mode from the infinite KK tower.
In the limit where $d \rightarrow \infty$, the expansion in
(\ref{expF}) is ill-defined as the function $f$ in (\ref{deff})
diverges and dominates over the zero-mode.
In the limit of one positive
brane embedded in a non-compactified fifth dimension
(RSII model), a derivative expansion is not possible. There is indeed no mass
gap between each mode in the KK tower when the fifth dimension is infinite.

To first order in derivatives, the perturbed metric induced on each
brane in its respective de Donder gauge is therefore:
\begin{eqnarray}
&& \bar{h}_{\mu \nu}^{(+)}=-\frac{2L  \kappa}{1-\Psi_0^2} \left[
\frac{1}{L^2 \Box}- f(0)
+\mathcal{O}(L^2 \Box)
\right]\, \Sigma_{\mu \nu}\label{huv+} \\
&&\phantom{\bar{h}_{\mu \nu}^{(+)}=}+\frac{\kappa}{3L}\eta_{\mu
  \nu}\frac{1}{\Box}T+\frac{2\kappa}{3 L}\frac{1}{\Box^2} T_{, \mu \nu},
\notag \\
&& \bar{h}_{\mu \nu}^{(-)}= -\frac{2L \kappa\,\Psi_0^2}{1-\Psi_0^2}
\left[ \frac{1}{L^2 \Box}- f(d)
+\mathcal{O}(L^2 \Box)
\right]\, \Sigma_{\mu \nu}. \label{huv-}
\end{eqnarray}


\subsection{Extension of the Standard Four-dimensional Effective
  Action}
\subsubsection{The Four-dimensional Effective Theory
\label{Section4deffaction}}
In this section, we suggest a possible extension of the
four-dimensional effective theory that is capable of recovering the first KK
mode as described in the previous section.

We work in a low-energy limit, when the density of the matter
confined on the branes is much smaller that the brane tensions. In
that case, it has been shown (see for example \cite{Mendes,KZ,Shiromizu}),
that the zero mode of the two brane Randall-Sundrum model can be
described by a four-dimensional effective theory.

In this theory, the metrics on both branes are conformally related:
\begin{eqnarray}
g_{\mu \nu}^{(-)}=\Psi^2 g_{\mu \nu}^{(+)}. \label{confrel}
\end{eqnarray}
This has already been mentioned in appendix \ref{section p2 terms},
and the
modified Einstein equation on the positive tension brane is:
\begin{eqnarray}
G^{(+)}_{\mu \nu}=\frac{\kappa}{L}\,T_{\mu \nu}
-E_{\mu \nu}^{(+)},
\end{eqnarray}
with $T_{\mu \nu}$ the stress-tensor of matter on the positive brane.
If we introduce matter only on this brane, the
expression for the tensor $E_{\mu \nu}^{(+)}$ in this effective theory is:
\begin{eqnarray}
E_{\mu \nu}^{(+)}=-
4 \partial_{\mu }\Psi
\partial_{\nu }\Psi
+2\Psi \nabla_{\mu} \nabla_{\nu} \Psi
-\left(2 \Psi \Box \Psi -\left(\partial \Psi \right)^2
\right)g_{\mu\nu}^{(+)}
-\Psi^2 G^{(+)}_{\mu \nu}, \label{Euv+}
\end{eqnarray}
where all covariant derivatives are taken with respect to
$g_{\mu\nu}^{(+)}$ (as will be the case throughout this section
unless otherwise specified).
This is in complete agreement
with the Gauss-Codacci equations in the low-energy limit
when the electric part of the induced Weyl tensor $E_{\mu \nu}^{(+)}$ is fixed to
the given value (\ref{Euv+}), (cf. appendix \ref{section p2 terms}).
Using the conformal transformation (\ref{confrel}),
this effective theory
predicts the electric part of the Weyl tensor on the negative tension
brane to be:
\begin{eqnarray}
E_{\mu \nu}^{(-)}=\Psi^{-2}\, E_{\mu \nu}^{(+)}.\label{Euv-}
\end{eqnarray}
These relations (\ref{Euv+},\ref{Euv-}) for $E_{\mu \nu}^{(\pm)}$
can be checked to be true for the background and
to model correctly the behaviour of the zero mode (ie. the long
wavelength limit of the exact five-dimensional theory). However there
is no reason for expressions (\ref{Euv+},\ref{Euv-})
 to remain exact for non-conformally
flat spacetimes in general.
The electric part of the Weyl tensor $E_{\mu \nu}^{(\pm)}$
is indeed the only quantity which remains unknown from a
purely four-dimensional point of view as it encodes information
from the bulk geometry. It is through this term that the bulk
generates KK corrections on the brane.

From the properties of the
five-dimensional Weyl tensor, it can be shown that
$E_{\mu \nu}^{(\pm)}$ is
traceless.
Furthermore, in the low-energy limit, by the Bianchi identity,
 it is divergenceless. (Outside the low-energy limit, as in the
 context of appendix \ref{section p2 terms}, $E_{\mu \nu}$
is not transverse but it is possible to separate out its transverse
part, noted as $E_{\mu
   \nu}^{TT}$.)
If we want to
modify the four-dimensional effective theory, the only modification
that would be consistent with the five-dimensional nature of the
theory
is to add a correction
 to the expression (\ref{Euv+}) of
$E_{\mu \nu}^{(+)}$, (or to (\ref{Euv-}) for $E_{\mu
   \nu}^{(-)}$). Motivated by \cite{toby,muk}, 
let us consider a possible term $E_{\mu \nu}^{corr}$ that could be added
 as a correction to (\ref{Euv+}). $E_{\mu \nu}^{corr}$ needs to vanish
 for conformal flat spacetimes since the relation (\ref{Euv+})
is exact in that case. Since $E_{\mu \nu}^{(+)}$ is transverse, so is
$E_{\mu \nu}^{corr}$, and we might think of it as being derived
 from an action $S_{E}$ (noting however that this is not
 necessarily the case).
Furthermore, since
$E_{\mu \nu}^{corr}$ is traceless,
$S_{E}$ must be conformally invariant. Indeed, the
variation of this action under a conformal transformation with
$\delta g_{\mu \nu} \propto g_{\mu \nu} $ will be:
\begin{eqnarray}
\delta S_{E}=\int d^4x \sqrt{-g}\frac{1}{2} E_{\mu \nu}^{corr} \delta
g^{\mu \nu},
\end{eqnarray}
and since $ E_{\mu \nu}^{corr}$ is traceless, the action $S_{E}$ must be
conformally invariant. \bigskip

If we want to modify the effective theory in order to accommodate the
first KK corrections, we need to add to the action a term of fourth order
in derivatives. If we consider a functional of the metric
only, the only possible local term at this order that preserves the
conformal invariance is:
\begin{eqnarray}
S_{C^2}=\int d^4x \sqrt{-g} \
C_{\alpha \beta \gamma \delta} C^{\alpha \beta \gamma \delta}.
\label{actionCC}
\end{eqnarray}
The stress-tensor derived from the action is:
\begin{align}
A_{\mu \nu }[g] =&\frac{1}{2\sqrt{-g}}
\frac{\delta\,  S_{C^2}}{\delta g^{\mu
\nu }}
,\notag  \\
A_{\mu \nu }[g] =&
-\nabla_{\alpha}\nabla^{\alpha} R_{\mu \nu}
+\frac{1}{3}\nabla_{\mu}\nabla_{\nu}R
+\frac{1}{6}\nabla_{\alpha}\nabla^{\alpha} R g_{\mu \nu}
-\frac{1}{6}R^2 g_{\mu \nu}  \label{Auv2}\\
& +\frac{2}{3}R R_{\mu \nu}
+\frac{1}{2}R_{\alpha \beta}R^{\alpha \beta} g_{\mu \nu}
-2 R_{\mu \alpha \nu \beta}R^{\alpha \beta}. \notag
\end{align}
Here all covariant derivatives are taken with respect to the general metric
$g_{\mu\nu}$.
Since the Weyl tensor vanishes for conformally flat spacetimes, the
tensor $A_{\mu \nu }[g]$ has all the requirements it
 should satisfy: it is indeed traceless, transverse and vanishes
for the background. Adding a term
$ E_{\mu \nu}^{corr}\propto A_{\mu \nu}[g^{(+)}]$ to the expression
(\ref{Euv+}) is therefore consistent with the Gauss-Codacci and the
Bianchi identities and is consistent with the background results.
Following this argument, it seems to us natural to consider the
effects that the addition of such terms would have in the
theory. \bigskip

Moreover, from the results of section \ref{sectionGarrigaTanaka}, 
we can point out
that when the first KK corrections are taken into account, the brane metrics
do not remain conformal to each other any longer.
We therefore need to
modify the relation (\ref{confrel}) between the brane metrics.

In our procedure we suggest breaking the conformal relation
 between the brane metrics by including some independent contribution
 to $E_{\mu \nu}^{(\pm)}$. By independent, we mean that the
 corrections will not satisfy the conformal relation (\ref{Euv-}),
 ie. 
$E_{\mu \nu}^{(-) corr}\ne \Psi^{-2} E_{\mu \nu}^{(+) corr}$.
%
%
\subsubsection{Ansatz}
Our idea is to include a contribution
 from the stress-tensor (\ref{Auv2})
in the effective theory. More
precisely, we will include a term proportional to $A_{\mu
 \nu}[g^{\pm}]$
to the
electric part of the Weyl tensor.
We will therefore consider the theory governed by the four-dimensional
equations:
\begin{align}
&
 G_{\mu \nu}^{(+)}=\frac{\kappa}{L}T_{\mu \nu}
+4 \partial_{\mu }\Psi
\partial_{\nu }\Psi
-2\Psi \nabla_{\mu} \nabla_{\nu} \Psi
+\left(2 \Psi \Box \Psi -\left(\partial \Psi \right)^2
\right)g_{\mu\nu}^{(+)}  \label{G+}\\
& \phantom{G_{\mu \nu}^{(+)}=}+\Psi^2 G^{(+)}_{\mu \nu}
-E_{\mu \nu}^{(+) corr }, \notag \\
&
G^{(-)}_{\mu \nu}=\frac{1}{\Psi^2}\left(
-2\Psi \nabla_{\mu}\nabla_{\nu}\Psi+4\partial_{\mu }\Psi
\partial_{\nu}\Psi+g^{(+)}_{\mu \nu} \left(2\Psi \Box \Psi
-\left|\nabla
\Psi \right| ^2 \right)
 \right)\label{G-}\\
& \phantom{G_{\mu \nu}^{(+)}=}
+G^{(+)}_{\mu \nu}-E_{\mu \nu}^{(-)corr},\notag \\
& \Box^{(+)} \Psi=\frac{1}{6}R^{(+)}\Psi,
\end{align}
with the corrections terms:
\begin{align}
& E_{\mu \nu}^{(+)corr}= \alpha\, L^2 A_{\mu \nu}[g^+],\label{Ecorr+} \\
& E_{\mu \nu}^{(-)corr}= \beta \, L^2 A_{\mu \nu}[g^-]. \label{Ecorr-}
\end{align}
We shall test this modified
four-dimensional effective theory against exact results obtained from
five-dimensional analysis in what follows.
%
%
\subsubsection{Verification of the Ansatz for Static Branes}
In this last section we shall check the previous ansatz against the exact
solution derived in section \ref{sectionGarrigaTanaka}, in the case of
perturbations around static branes. The ansatz we propose has to be
considered as a first order correction and we will check if it
correctly reproduces the behaviour of the first KK mode from
the previous section. We will therefore not solve the system
completely but only work out the zero mode and the
first order correction generated by the
additional terms in (\ref{Ecorr+},\ref{Ecorr-}).
We study the same scenario as in section \ref{sectionGarrigaTanaka}
and we consider metric perturbations around static
branes, sourced by the addition of
matter on the positive tension brane. In the background,
the positive tension brane is located at $y=0$ and the negative one at
$y=d$.
To first order in perturbations,
\begin{eqnarray}
&& g^{(+)}_{\mu \nu}=\eta_{\mu \nu}+\bar{h}^{(+)}_{\mu \nu}, \\
&& \Psi=\Psi_{0}+\delta \Psi, \ \ \Psi_{0}=e^{-d/L}=\text{const.},\\
&& g^{(-)}_{\mu \nu}=\Psi_{0}^2 \ \eta_{\mu \nu}+\bar{h}^{(-)}_{\mu
\nu}.
\end{eqnarray}
To first order in perturbations,
we can consider the matter fields and the perturbed
scalar field $\delta \Psi$
to live on the background metric:
\begin{eqnarray}
&& R^{(+)}=-\frac{\kappa}{L}T,\\
&&\Box \delta \Psi=-\frac{\kappa}{6 L}T\Psi_0, \\
&& A_{\mu \nu}^{(+)}=-\Box R_{\mu \nu}^{(+)}+\frac{1}{3} R^{(+)}_{,\, \mu \nu}
+\frac{1}{6}\Box R^{(+)} \eta_{\mu \nu},
\end{eqnarray}
giving rise to the modified Einstein equation for the positive tension
brane:
\begin{eqnarray}
&& R_{\mu \nu}^{(+)}= \frac{\kappa/L}{1-\Psi_0^2}\left[
T_{\mu \nu}-\frac{1}{2}T \eta_{\mu \nu}
+\frac{\Psi_{0}^2}{6}
T\eta_{\mu \nu}
+\frac{\Psi_0^2}{3\,  \Box}T_{, \mu \nu}\right] \\
&& \phantom{R_{\mu \nu}^{(+)}}
-\frac{\alpha \kappa /L }{(1-\Psi_0^2)^2}\, L^2 \Box
\left[
T_{\mu \nu}-\frac{1}{3}T \eta_{\mu \nu}
+\frac{1}{3\, \Box}T_{, \mu \nu}
\right]\notag \\
&& \phantom{R_{\mu \nu}^{(+)}} \notag
+\frac{\kappa}{L}\ \mathcal{O}\left(\alpha ^2 L^4 \Box^2 T_{\mu \nu}\right).
 \end{eqnarray}
To get this result, we have only considered the first order
corrections generated by the addition of the term (\ref{actionCC}) in
the effective four-dimensional action. We notice that up to first
order in perturbations and to first order in the
 correction term $\alpha L^2 \Box$,
we have the remarkable relation:
\begin{eqnarray}
A_{\mu \nu}^{(+)}=-\frac{ \kappa /L }{1-\Psi_0^2}\, \Box
\Sigma_{\mu \nu},
 \label{Auv_Sigma}
\end{eqnarray}
with the tensor $\Sigma_{\mu \nu}$ as given in (\ref{defSigma}).
This is an important result, based on which, our ansatz will be verified.

For the negative tension brane, the modified Einstein equation reads:
\begin{eqnarray}
R_{\mu \nu}^{(-)}=R_{\mu \nu}^{(+)}-2\frac{\delta \Psi_{,\, \mu
    \nu}}{\Psi_0}
-\frac{\Box \delta \Psi}{\Psi_0}\eta _{\mu \nu}-\beta L^2 A_{\mu \nu}^{(-)},
\end{eqnarray}
so that:
\begin{eqnarray}
R_{\mu \nu}^{(-)}=
\frac{\kappa/L}{1-\Psi_0^2}
\left[
1-\left(\beta
(1-\frac{1}{\Psi_0^2})+\frac{\alpha}{1-\Psi_0^2}\right)L^2 \Box
\right]\Sigma_{\mu \nu}.
\end{eqnarray}
In the de Donder gauge, the Ricci tensor is related to the
metric perturbation by: $R_{\mu \nu}^{(\pm)}=
-\frac{1}{2}\Box^{(\pm)} \bar{h}^{(\pm)}_{\mu \nu}$. 
The perturbation of the brane metrics is
therefore:
\begin{eqnarray}
&& \bar h_{\mu \nu}^{(+)}=
- 2\frac{\kappa L}{1-\Psi_0^2} \left[\frac{1}{L^2 \Box}
-\frac{\alpha}{1-\Psi_0^2}+\mathcal{O}(L^2 \Box)
\right]\Sigma_{\mu \nu}\label{h++}\\
&& \phantom{\bar h_{\mu \nu}^{(+)}=}
+\frac{2\kappa}{3L\Box^2}T_{,\, \mu \nu}+\frac{\kappa}{3 L \Box}T \eta_{\mu
\nu}, \notag \\
&& \bar h_{\mu \nu}^{(-)}=-2
\frac{\kappa L \Psi_0^2}{1-\Psi_0^2}
\left[
\frac{1}{L^2 \Box}-\left(\beta (1-\frac{1}{\Psi_0^2})
+\frac{\alpha}{1-\Psi_0^2}\right)
\right]\Sigma_{\mu \nu}.\label{h-+}
\end{eqnarray}
We can compare these results with the exact zero and first
modes obtained in (\ref{huv+}) and (\ref{huv-}). We can check that the zero modes
agree perfectly as one should expect. This just confirms the
well-known result that the effective theory gives the correct zero
mode on the brane.

 More remarkable is the result
for the first KK mode. Comparing the expression (\ref{h++})
with (\ref{huv+}), the
results agree perfectly if the dimensionless constant $\alpha$ is
fixed to the value: $\alpha=(1-\Psi_0^2)f(0)$. Similarly, we have a
perfect agreement between the ansatz for
the negative tension brane (\ref{h-+}) and the exact result
(\ref{huv-}) if the constant $\beta$ is fixed to the value:
$\beta=\frac{\Psi_0^2}{1-\Psi_0^2}\left(f(0)-f(d)\right)$. \bigskip

This is an original result. It has been possible to extend the
notion of a four-dimensional effective theory in order to
accommodate the first KK correction arising from perturbations
around static branes. \bigskip

We should however note that in order to fit to the exact results,
the coefficients $\alpha$ and $\beta$ have to depend on the
distance between the branes, or equivalently depend on the scalar
field $\Psi$. If the scalar field is not fixed in the background,
$\partial_{\mu}\Psi_0 \ne 0$, in general the action
$\int d^4x \sqrt{-g} \, \alpha(\Psi)
C_{\alpha \beta \gamma \delta}C^{\alpha \beta \gamma \delta}$ is not
conformally invariant, and our procedure is not valid.
The generalisation of the effective theory to the case
for which the branes are not fixed in the background
is therefore not straightforward. In that case a more general analysis
is needed as in \cite{KS}. However, to our knowledge, an exact
five-dimensional derivation for the KK modes in a more general
scenario has not yet been studied. It is therefore difficult to check
any new ansatz for a four-dimensional effective theory in any more
elaborate scenario. We hope that our work can be used in order to check
the validity of a four-dimensional effective theory and gives
insight on how to generalise it to take some typical five-dimensional
features into account.


\end{document}